\documentclass[a4paper]{article}

\usepackage{hyperref}
\usepackage{amsmath,amsfonts,amssymb}
\usepackage{amsthm}
\usepackage{graphicx}
\usepackage{lscape,epsf}
\usepackage[vcentermath]{youngtab}
\usepackage{rotating}
\usepackage[usenames,dvipsnames]{xcolor}

\def\ds{\displaystyle}
\def\bea{\begin{array}{c}}
\def\ea{\end{array}}
\def\be{\begin{equation}\bea\ds}
\def\ee{\ea\end{equation}}
\def\bee{\begin{equation}\begin{array}{rcl}\ds}
\def\eee{\end{array}\end{equation}}
\DeclareMathOperator{\sign}{sign}

\def\nb{{\bf n}}
\def\bb{{\bf b}}
\def\tb{{\bf t}}

\def\xb{{\bf x}}

\def\ub{{\bf u}}

\def\htau{\hat{\tau}}
\def\hk{\hat{\kappa}}

\def\eb[#1]{{\bf e}_{#1}}
\def\heb[#1]{\hat{\bf e}_{#1}}

\title{\bf Topological Indices of Proteins}

\author{Dmitry Melnikov$^{a,b}$, Antti J.~Niemi$^{c,d,e}$ and Ara Sedrakyan$^{a,f}$}

\date{}

\begin{document}

\maketitle

\vspace{-5cm}
\hfill{ITEP-TH-22/18}
\vspace{5cm}

\vspace{6pt}
\begin{center}
\textit{\small $^a$ International Institute of Physics, Federal University of Rio Grande do Norte, \\ Campus Universit\'ario, Lagoa Nova, Natal-RN  59078-970, Brazil}
\\ \vspace{6pt}
\textit{\small $^b$  Institute for Theoretical and Experimental Physics, B.~Cheremushkinskaya 25, Moscow 117218, Russia}\\ \vspace{6pt}
\textit{\small $^c$  Nordita, Stockholm University, Roslagstullsbacken 23, SE-106 91 Stockholm, Sweden}\\ \vspace{6pt}
\textit{\small $^d$  Laboratory of Physics of Living Matter, Far Eastern Federal University,\\ Sukhanova 8, Vladivostok 690950, Russia}
\\ \vspace{6pt}
\textit{\small $^e$  Department of Physics, Beijing Institute of Technology,
Haidian District, Beijing 100081, P. R. China}\\ \vspace{6pt}
\textit{\small $^f$  National Laboratory after Alikhanyan, Yerevan Physics Institute, Br.Alikhanyan 2, Yerevan 0036, Armenia}
\\ \vspace{2cm}
\end{center}

\begin{abstract}
Protein molecules can be approximated by discrete polygonal chains of amino acids. Standard topological tools can be applied to the smoothening of the polygons to introduce a topological classification of proteins, for example, using the self-linking number of the corresponding framed curves. In this paper we add new details to the standard classification. Known definitions of the self-linking number apply to non-singular framings: for example, the Frenet framing cannot be used if the curve has inflection points. Meanwhile in the discrete proteins the special points are naturally resolved. Consequently, a separate integer topological characteristics can be introduced, which takes into account the intrinsic features of the special points. For large number of proteins we compute integer topological indices associated with the singularities of the Frenet framing. We show how a version of the Calugareanu's theorem is satisfied for the associated self-linking number of a discrete curve. Since the singularities of the Frenet framing correspond to the structural motifs of proteins, we propose topological indices as a technical tool for the description of the folding dynamics of proteins.
\end{abstract}

\newpage

\section{Introduction}


Up to space translations and rigid rotations, curves in three dimensions can be defined in terms of a pair of scalar functions of a single scalar parameter. One possible choice is curvature $\kappa(s)$ and torsion $\tau(s)$ (here selected to be functions of the arc-length parameter $s$), which respectively provide a local measure of the given curve failing to be straight and planar. Curvature and torsion characterize the local rotation of a right triple of vectors (Frenet frame) $\{\nb(s),\bb(s),\tb(s)\}$ along the curve. Here $\tb$ is the tangent vector, $\nb$ and $\bb$ are the normal and binormal vectors respectively. Given functions~$\kappa(s)$ and $\tau(s)$ one can recover the frame at each point and the parameterization $\xb(s)$, up to space isometries, using for example, Frenet equations~\cite{Frenet:1852}. 

\begin{figure}[t]
 \begin{minipage}{0.5\linewidth}
 \includegraphics[width=\linewidth]{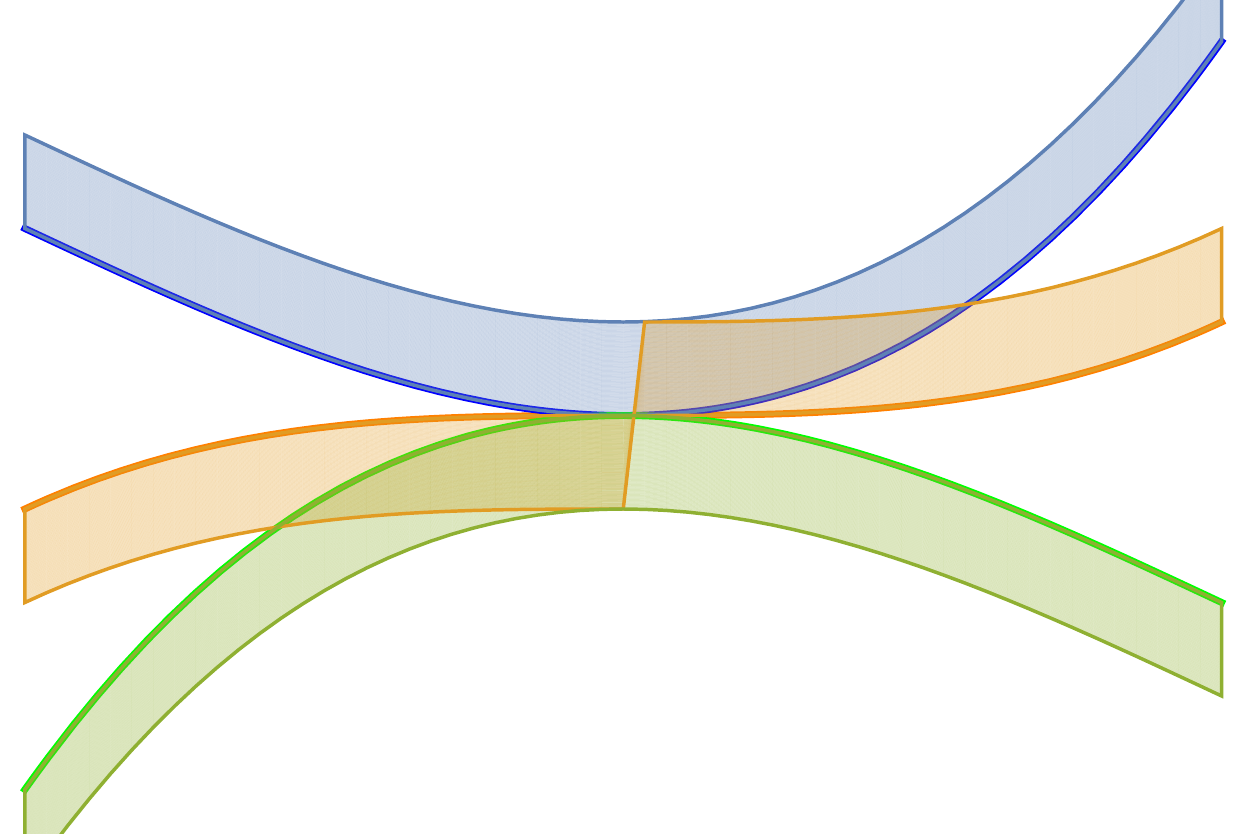}
\end{minipage}
\hfill{
 \begin{minipage}{0.4\linewidth}
\centering
 \includegraphics[width=0.45\linewidth]{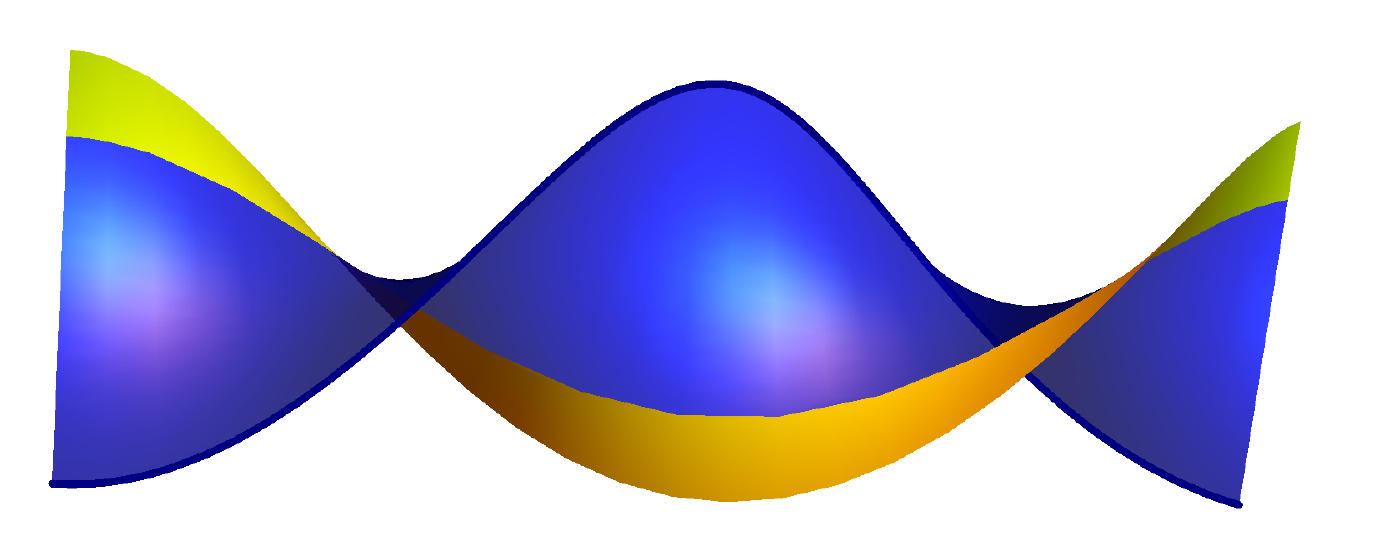}
 \hfill{
\includegraphics[width=0.45\linewidth]{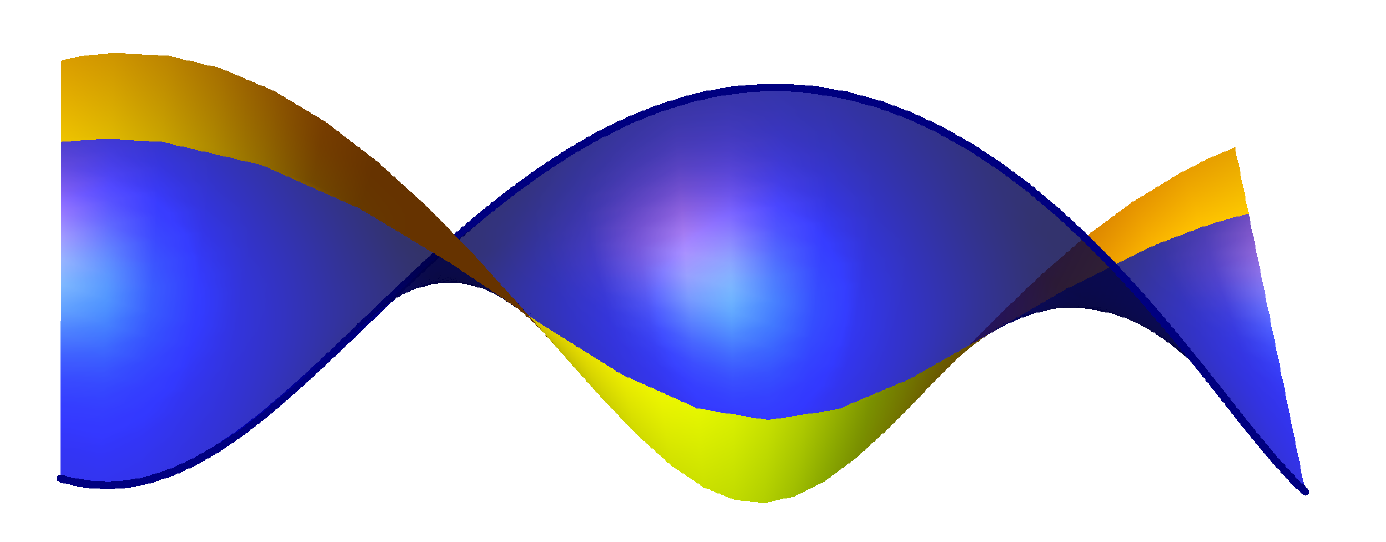}
} \\
 \includegraphics[width=0.45\linewidth]{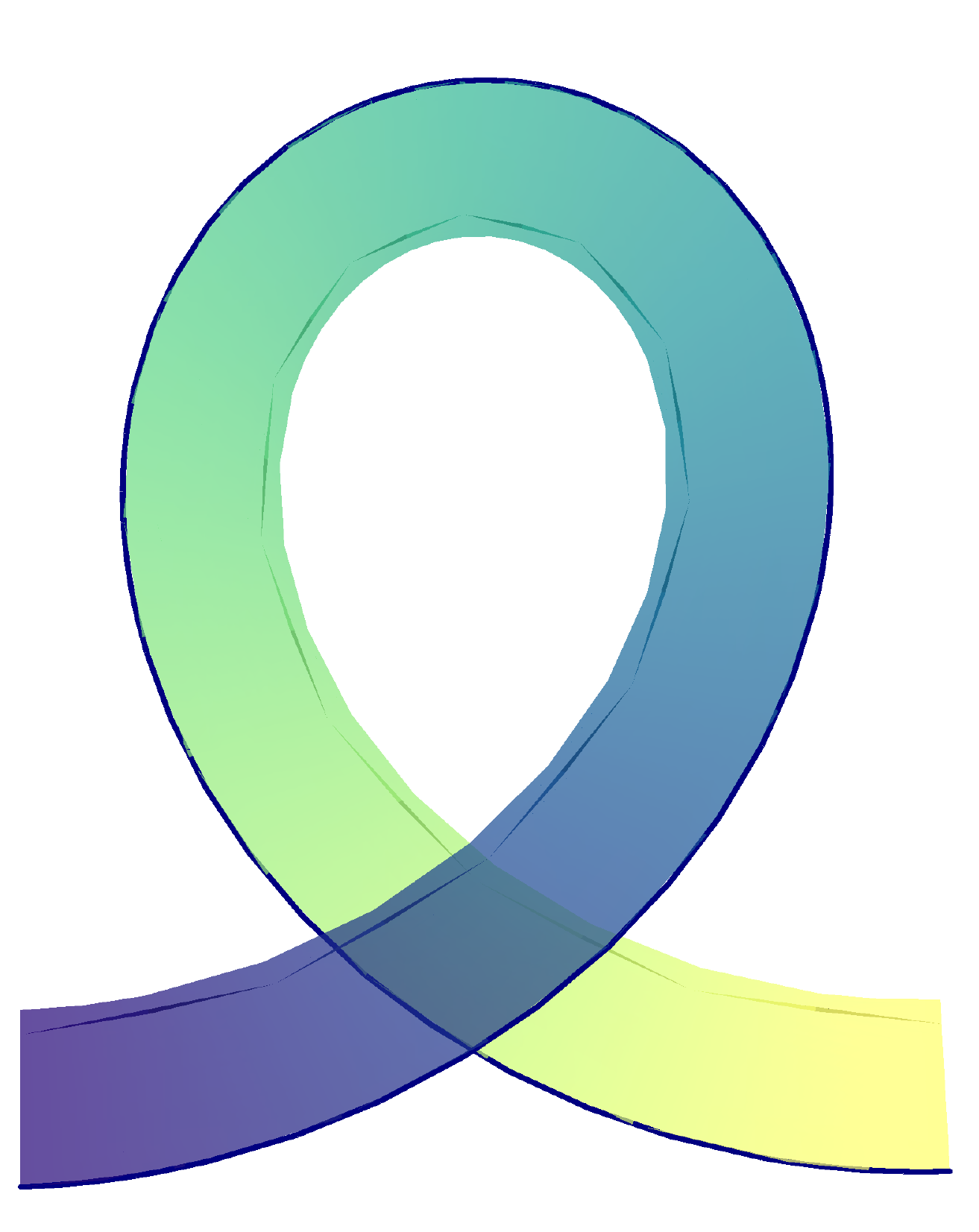}
 \hfill{
\includegraphics[width=0.45\linewidth]{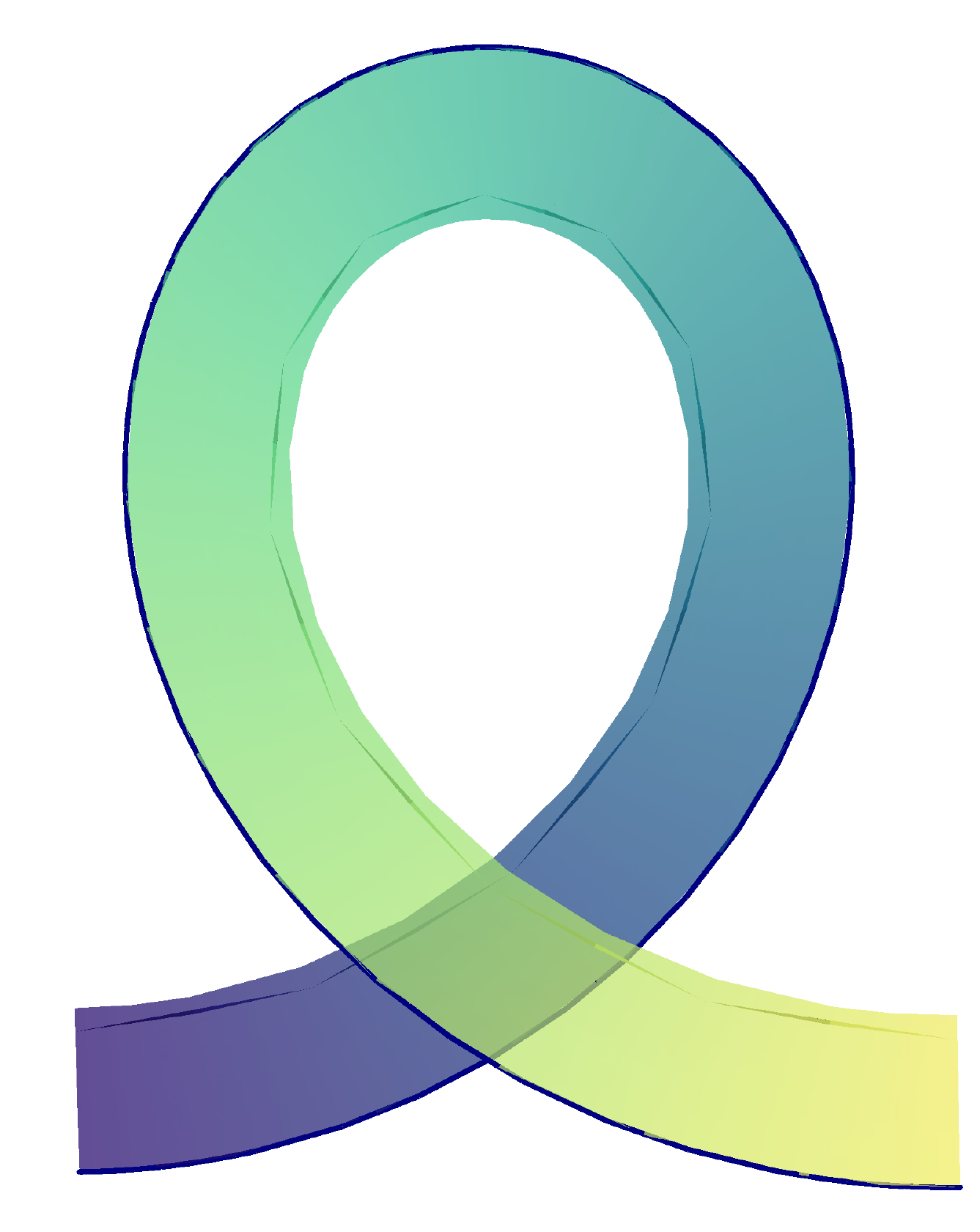}
}
\end{minipage}
}
 \caption{(Left) Three members of a one-parametric family of planar curves corresponding to positive curvature (blue), negative curvature (green) and curvature that changes sign in the middle point (orange). The Frenet framing of the orange curve has a discontinuous jump of the direction of the normal vector across the inflection point. (Right) Segments of a framed 3D curve with $tw=-1$ (top left), $tw=1$ (top right), $wr=-1$ (bottom left) and $wr=1$ (bottom right).}
 \label{fig:inflection}
\end{figure}

Given curvature and torsion, Frenet equations define a \emph{framed} curve, that is a ribbon defined by tangent vector $\tb$ and a vector $\ub$ transverse to $\tb$ at every point on the curve. The Frenet frame corresponds to a particular choice $\ub=\bb$ (or $\ub=\nb$). It is not always a convenient choice, since at \emph{inflection points}, where $\kappa=0$, the direction of $\bb$ and $\nb$ is not defined. These vectors experience a discrete jump by angle $\pi$ across the inflection point (see the left panel of figure~\ref{fig:inflection}). Nevertheless, for a smooth curve, one can always introduce a different framing, well-defined at inflection points.

Framed curves can be endowed with a topological characteristic called \emph{self-linking number} $Lk$, introduced by the Gauss integral formula
\be
Lk \ = \ \frac{1}{4\pi} \oint_\gamma ds \oint_{\gamma_{\bf u}}ds' \ \frac{[\dot{\xb}(s)\times \dot{\xb}(s')]\cdot(\xb(s)-\xb(s'))}{|\xb(s)-\xb(s')|^3}\,, 
\ee 
where $\gamma$ is a closed curve and $\gamma_{\ub}$ is its framing, \emph{i.e.} a curve parameterized by $\xb(s)+\epsilon\ub(s)$, with some small $\epsilon>0$. The integral computes the Gauss' linking number of $\gamma$ and $\gamma_{\ub}$. It is an integer number invariant under smooth generic deformations of the curve, or its framing. Provided the above mentioned properties of the Frenet framing, one should conclude that it is not a good choice for the calculation of the self-linking number of a curve with inflection points~\cite{Aicardi}. Nevertheless, the Frenet framing can be useful in detecting such points, providing an additional information about the curve. As non-generic points of the three-dimensional embedding of a segment (or a circle), inflection points can have a physical significance. 

The purpose of this paper is to show how one can use the Frenet framing and some of its extensions to detect and classify special points of the discretized versions of curves, the polygons. The primary motivation of this exercise is to connect the special points of framed polygons with the secondary structure motifs in protein molecules and understand their role in the folding process as well as the biological function of the proteins.

We start from a theorem of Calugareanu~\cite{Calugareanu} about basic topology of closed curves. Self-linking number can be calculated from a two-dimensional diagram of a framed curve, obtained by its projection on a selected plane. Calugareanu's theorem (Calugareanu-White-Fuller~\cite{White,Fuller}) states that the self-linking number is provided by the sum,
\be
\label{Calugareanu}
Lk \ = \ wr + tw\,,
\ee
of two quantities known as writhe and twist. The writhe is defined as a difference of positive and negative self-intersections of the projection $P(\gamma)$ of the curve $\gamma$ on the plane (bottom right of figure~\ref{fig:inflection}), while the twist is the half-difference of positive and negative intersections of $P(\gamma)$ and $P(\gamma_{\ub})$ (top right of the same figure). Under the smooth variations (isotopies) of the three-dimensional curve, or under the change of the projection plane, writhe and twist can only change in such a way that their sum $Lk$ is preserved.\footnote{See~\cite{Kauffman:2001,Dennis} for three-dimensional definitions of $tw$ and $wr$.}

Calugareanu's formula can be applied to examples of extended quasi-one-dimensional objects in biophysics and soft matter, for example in the DNA supercoiling and folding of polymers and protein molecules. See~\cite{Fuller,BioTopology} for references. A common point of those studies can be put as follows: topology, to a certain extent, controls dynamics of such biological systems.

In this paper we discuss a special version of the self-linking number applicable to protein molecules. This number arises as a result of a ``resolution" of the ill-defined linking number associated with the Frenet framing. For proteins, which are not closed curves, the total self-linking number can be defined as a sum of a regular, but not integer, piece, computed with respect to a non-singular framing, and an integer piece, characterizing the singular points of the Frenet framing,
\be
Lk \ = \ Lk_0 + \frac{\theta}{2}\,, \qquad \theta\in \mathbb{Z}\,.
\ee
We argue that in the Frenet framing index $\theta$ has a meaning of the twist, counting large local rotations of the normal vector. One can locally undo the large rotations by a transformation, which do not change the curve, but introduces a different framing (gauge transformation). In the new framing the total self-linking number can be written as
\be
Lk \ = \ Lk_0 + \frac{\omega}{2}\,, \qquad \omega\in \mathbb{Z}\,,
\ee
where $\omega$ has the meaning of writhe.

Similarly to the Calugareanu's theorem, one would like to claim that for a generic framing there is an invariant given by the sum $\omega+\theta$. There is however a complication related to the fact that the Frenet framing is singular at the inflection points. In particular, the signs of the discrete contributions to $\omega$ and $\theta$ are not uniquely defined. We propose that the correct choice for $\omega$ and $\theta$ is the one, which simply counts the number of singularities of the Frenet framing. This number is a topological invariant, satisfying a version of the Calugareanu's theorem.

The paper is organized as follows. In section~\ref{sec:matrix} we give a definition of the self-linking number in terms of the matrices rotating frames along the curves. The version of this definition applicable to discrete curves is discussed in section~\ref{sec:discrete}. In section~\ref{sec:indices} we introduce and compute indices for discrete curves, which are shown to take almost integer values for protein molecules. In particular, we compute two indices for two different choices of discrete framing. In section~\ref{sec:main} we explain how the integer indices are related to the self-linking number, as defined in section~\ref{sec:discrete}. We conclude that the two indices are similar to the twist and writhe. Specifically, Frenet framing is a special framing with a pure twist index. We observe that the two indices, and the corresponding twist and writhe, are not equal for the framing choices naturally provided by the proteins. We explain how the definition of the twist and writhe must be rectified in order be compatible with the Calugareanu's theorem. In section~\ref{sec:gauge} we briefly discuss the continuous presentation of curves and the associated gauge symmetry. Loci of the indices are solitons of this continuous model. We conclude in section~\ref{sec:conclusions}.


\section{Matrix Presentation}
\label{sec:matrix}

It is convenient to use a different way to compute the self-linking number, which stems from the gauge theory formulation of topological invariants of knots and links~\cite{Polyakov}. This definition is obtained as follows. The frame $\eb[a]\equiv \{\nb(s),\bb(s),\tb(s)\}$ can be conveniently presented as a $2\times 2$ matrix using Pauli matrices $\hat\sigma^{i}$, $i=1,2,3$. In other words, we would like to work with spinor representation of the frame vectors.\footnote{In the topological context such presentation of strings was introduced in~\cite{Kavalov:1987vm,Kavalov:1986xw}. The spinor representation of Frenet equation was also considered in~\cite{Ioannidou:2014bxa}.} Matrix $\hat{e}$ representing one of the three frame vectors can be introduced as a contraction
\be
\hat{e}_a \equiv \{\hat{n},\hat{b}, \hat{t}\}\ = \ \sum\limits_i e_a^i \hat\sigma^i\,.
\ee
Consequently, three-dimensional rotations will be acting on $\hat{e}$ by $SU(2)$ matrices, such that rotation about axis $\ub=\{u^i\}$ by angle $\alpha$ is realized by a matrix 
\be
S \ = \ \exp\left(i\,\frac{\alpha}{2}\,\hat{u}\right)\,, \qquad \hat{u} \ = \ \sum\limits_i  u^i\hat\sigma^i\,.
\ee
Note that alternatively one can define vector matrices $\heb[]=\eb[a]\cdot\hat{\sigma}^a$. In such notations, the frames will be rotated by matrices acting in a different space, on indices of a different type. However, the two approaches are equivalent, the difference between them is such that, in the first case, one rotates the frame vectors with respect to a fixed choice of the basis vectors in space, while in the second case, the basis vectors are rotated with respect to a fixed-position frame. 

The frame at point $s$ is defined as a $SU(2)$ rotation $S(s)$ of the frame at the origin: 
\be
\label{Frenet2}
\hat{e}_a(s) \ = \ S(s)\hat{e}_a(0)S^{-1}(s)\,, \qquad a\ = \ 1,2,3\,.
\ee
We note that Frenet equations are just the infinitesimal form of the above formula, while $\kappa(s)ds$ and $\tau(s)ds$ are local infinitesimal rotation angles. Given $\kappa(s)$ and $\tau(s)$ one can recover the rotation matrix at any point $s$ in terms of a path ordered exponential
\be
\label{Pexponent}
S(s) \ = \ P\exp\left(-\int_{\gamma_s} SdS^{-1} \right)\,, \qquad SdS^{-1}\ = \ \frac{i}2\left(
\begin{array}{cc}
\tau & -i\kappa \\
i\kappa & -\tau
\end{array}
\right)ds,
\ee
where $\gamma_s$ is a part of the curve parameterized by the arc-length parameter on the interval~$[0,s]$.

If the curve is closed the frame should return to itself after completing a closed path along it. In this case $S(0)S^{-1}(L)=\pm I$, where $L$ is the length of the curve. We note that in Eq.~(\ref{Pexponent}) the integrand is a flat $SU(2)$ connection, characterized by an integer number of possible singular points of the non-Abelian field strength of the gauge field. Consequently the integral over a closed curve is proportional to an integer number times $\pi$. This integer number characterizing the total rotation of the frame along the curve is the self-linking $Lk$, $S=\exp(\pi i\cdot Lk)$. It can also be related to homotopy classes $\pi_3(S^2)$ of the Hopf map $S^3 \rightarrow S^2$~\cite{Kavalov:1987vm,Kavalov:1986xw}.

For open curves number $Lk$, computed by integral~(\ref{Pexponent}), is not an integer. In general, it describes the rotation between the initial and the final orientation of the frame. Since the space of such rotations is not simply connected, the latter split into non-trivial equivalence classes, distinguished by integer numbers, which are equivalent to $Lk$ for closed curves. 


\section{Geometry of discrete chains}
\label{sec:discrete}

Now we would like to discuss how the above definitions work in the case of discretized curves, \emph{i.e.} polygonal chains. Moreover, we will focus on a nature-given set of chains, the proteins. Protein molecules are quasi-one-dimensional sequences of amino acids. To visualize their geometric structure and study the folding dynamics, a coarse grained description of the molecule is commonly used. One choice is to represent the chain by the positions of the $C_\alpha$ carbon atoms of the amino acids. For our purposes $C_\alpha$ chain will be a discretized version of a smooth three-dimensional curve. Note that proteins provide a specific class of polygonal chains: chemical bonds and steric (excluded volume) constraints introduce a large degree of regularity to the protein molecules, as will be reviewed below. In particular, this allows to describe proteins in terms of a spin-chain-like model,~\emph{cf.}~\cite{Bryngelson:1987}. We will discuss the consequences of the constraints for topology.

\begin{figure}[t]
\begin{minipage}{0.7\linewidth}
 \includegraphics[width=\linewidth,clip=true,trim=50pt 400pt 100pt 450pt]{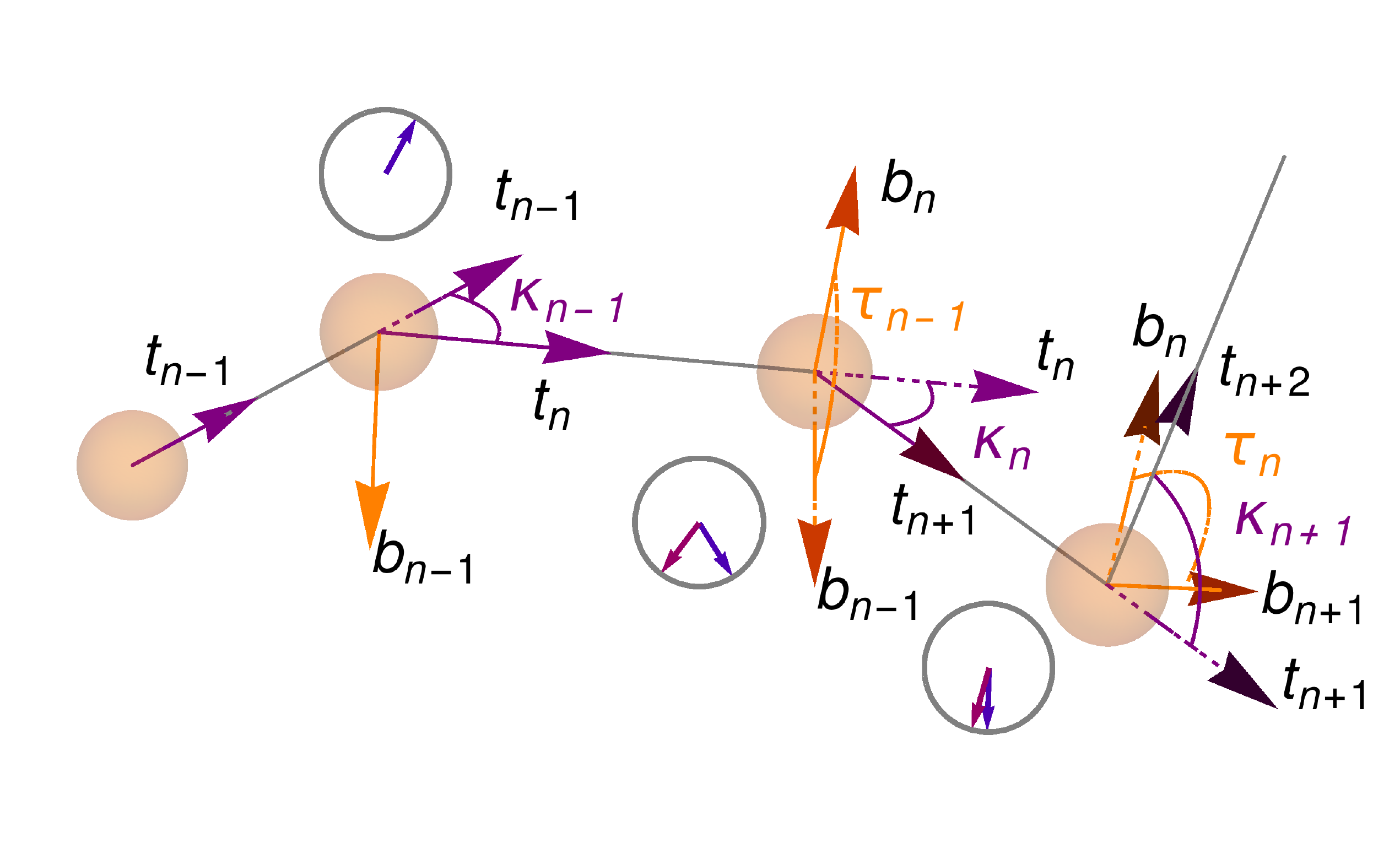}
\end{minipage}
\hfill{
\begin{minipage}{0.2\linewidth}
\centering
\includegraphics[width=\linewidth]{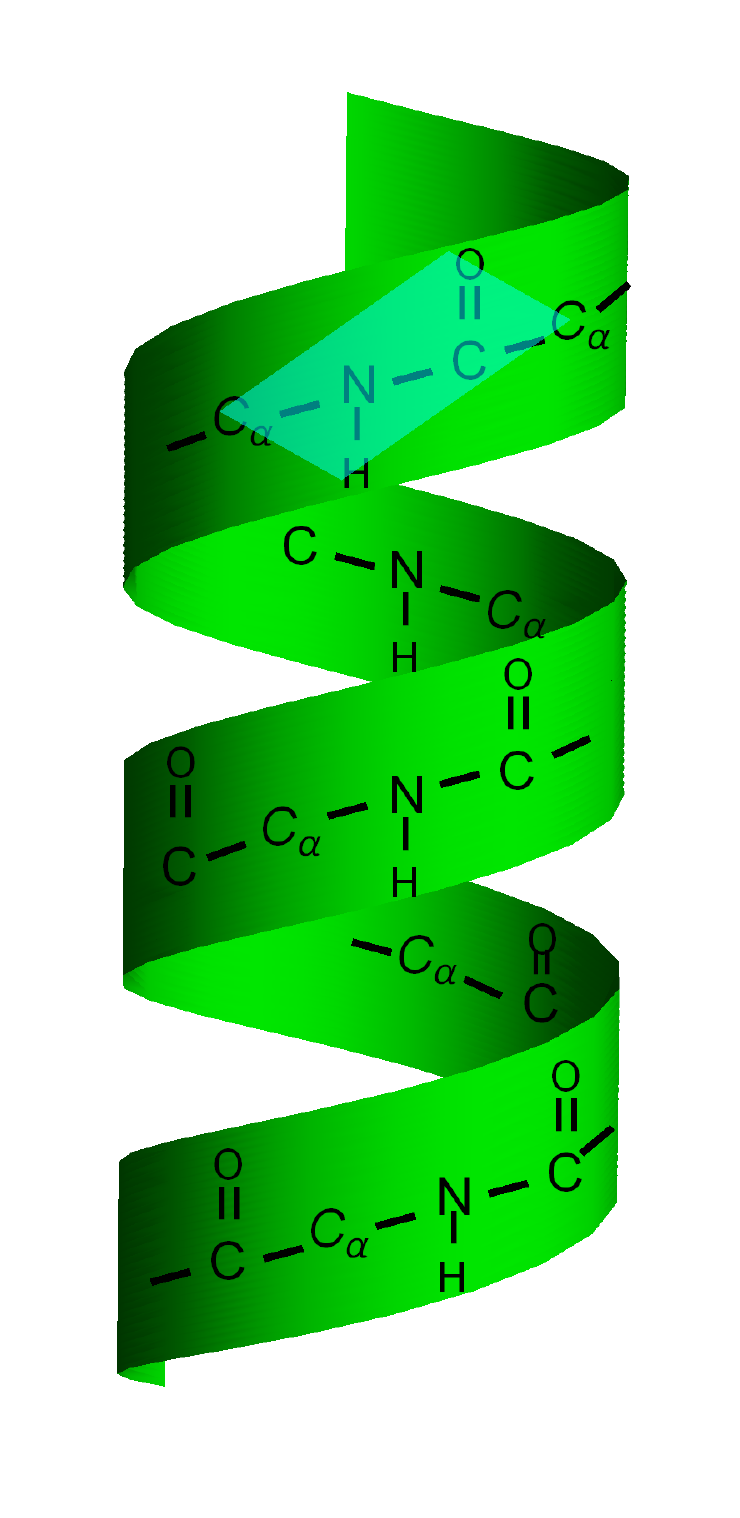}
\end{minipage}
}
 \caption{(Left) Definition of the geometric parameters. Curvature angles $\kappa_n$ define rotation of vectors $\tb_n$ towards adjacent $\tb_{n+1}$ around $\bb_n$. Torsion angles $\tau_n$ define rotations of binormal vectors $\bb_n$ towards the subsequent vector $\bb_{n+1}$ around $\tb_{n+1}$. Clocks show the magnitude of the angles for either curvature (blue) or torsion (purple). (Right) Illustration showing peptide framing of the protein chain. In particular, the bonds connecting $C_\alpha$ and side chains are not shown, because they do not belong to the peptide planes. The parallelogram illustrates a single peptide plane.}
  \label{fig:vectors}
\end{figure}

First, we introduce the discretized Frenet framing (left panel of figure~\ref{fig:vectors}). The nodes of the polygon $\xb_n$ are labeled by index $n$ running from one to $N$. Tangent vectors are defined as normalized differences of the positions of the $C_\alpha$ atoms, $ \tb_n \propto \xb_{n+1}- \xb_n$. Binormal vectors are normal to the plane of the local rotation of the tangent vectors,  $\bb_n \propto \tb_n\times \tb_{n+1}$, while $\nb_n$ complete the right triples (see~\cite{Chernodub:2010xz,Hu:2011wg} for more details).

Using the discrete Frenet frames one can define curvature and torsion angles, $\kappa_n$ and $\tau_n$, which serve as discrete versions of the continuous curvature $\kappa(s)$ and torsion $\tau(s)$. As can be deduced from the infinitesimal form of the rotation matrix in equation~(\ref{Pexponent}) $\kappa_n$ is the infinitesimal rotation (bond) angle around the binormal vector $\bb_n$, while $\tau_n$ is the infinitesimal rotation (twist) angle around the corresponding tangent vector. From an example on figure~\ref{fig:vectors} (left) one can calculate the discrete rotation angles as follows.
\begin{eqnarray}
\kappa_n & = &  \arccos(\tb_n\cdot \tb_{n+1})\,, \qquad n=1,\ldots,N-2\,, \label{dcurvature}\\
\tau_n & = & s_n \arccos(\bb_{n}\cdot \bb_{n+1})\,, \qquad n=1,\ldots,N-3. \label{dtorsion}
\end{eqnarray}
Here $0\leq \kappa_n<\pi$, while $\tau_n$ is defined with the sign $s_n=\pm1$, which determines the direction of the rotation of the frame around the tangent vector. This direction can be determined using two consecutive binormal vectors $\bb_{n}$ and $\bb_{n+1}$ connected by the bond parallel to $\tb_{n+1}$ using equation
\be
\label{sign}
s_n  \ = \  \sign ([\bb_{n+1}\times \bb_n]\cdot\tb_{n+1})\,.
\ee  
Consequently, $-\pi\leq \tau_n<\pi$. Note that torsion angles are defined with respect to two consecutive frames, which are subject to the condition that binormal vectors of both frames are both perpendicular to the bond connecting the frames and to the respective tangent vector. The curvature angle is only defined locally, for a single frame. The direction of the rotation around the binormal vectors is fixed and the angle takes values in $[0,\pi]$.

An alternative ``natural" framing is provided by the peptide planes of the protein chain. The peptide planes can be defined by the subsequent $C_\alpha C$ and $CN$ bonds along the backbone chain. A non-trivial fact is that also the oxygen of the $CO$ bond and the hydrogen of the $NH$ bond, as well the next $C_\alpha$ connected to $N$, all lie approximately within the same plane (there are six atoms contained in one plane). At the level of secondary structures, the peptide chain is traditionally visualized as a ribbon, which can be thought as a smoothening of the sequence of the connected peptide planes (approximately as on the right of figure~\ref{fig:vectors}). The ribbon version of the chain introduces a \emph{peptide framing} of the protein molecule.

Instead of the natural peptide framing, in this paper we will focus on the study of the discrete Frenet framing. We will show that this framing has special points and that those points correspond to secondary structure motifs. To that end we will need to compute torsion and curvature angles and the self-linking numbers of a series of proteins. Following the above discussion the self-linking number is computed by the discrete version of path ordered exponential~(\ref{Pexponent}), which is simply a product of discrete rotation matrices ordered along the chain:
\be
\label{DiscretePexp}
S(N) \ = \ S_{N,N-1}\cdot S_{N-1,N-2}\cdots S_{3,2}\cdot S_{2,1}\,.
\ee

According to figure~\ref{fig:vectors}~(left) there is a natural parametrization of the rotation matrices at each step: first, one rotates the frame around vector $\bb_n$ by angle $\kappa_n$, and next, rotates around $\tb_{n+1}$ by angle $\tau_n$,
\be
\label{elemrot}
S_{n+1,n} \ = \ {\rm e}^{i\frac{\tau_n}{2}\,\hat{t}_{n+1}}\cdot {\rm e}^{i\frac{\kappa_n}{2}\,\hat{b}_n}\,.
\ee
For a closed curve, the result should just be positive or negative identity matrix, $S=\pm I=\exp(\pi i Lk)$, where an integer self-linking number appears as an ambiguously defined phase.

We also note an ambiguity in the definition of the sign of the torsion angle in equation~(\ref{sign}) if $\bb_{n+1}=\pm\bb_{n}$.  The case $\bb_{n+1}=\bb_n$ is a flattening point, torsion angle vanishes across the bond, but the curvature is well defined so there is no actually an ambiguity. The case $\bb_{n+1}= - \bb_{n}$ is analogous to an inflection, which is accompanied by a $\pi$ flip of the direction of the normal and binormal vectors. 

Our main interest will be in the latter special points, at which the binormal vector changes sign. If the flip is an exact reflection by $\pi$, then one cannot say, whether this gives a positive, or a negative contribution to the self-linking number. On the other hand, in a nature-given chain like protein, the rotation is never exactly by $\pi$, so modulo possible experimental error, or even larger rotations by multiples of $2\pi$, the direction of the rotation can be defined unambiguously.


\section{Indices of Proteins}
\label{sec:indices}

We will now analyze the topological data of a set of proteins, whose structure was obtained with higher resolution and passed additional consistency checks.\footnote{Technically, these are PDB structures with homology equivalence less than 30\%, which  are obtained using diffraction data with the resolution better than 1.0{\AA} and verified to only include structures, which do not have a unit cell containing more than one peptide chain, which do not have a missing heavy atom in the backbone, or do not have alternate positions for heavy atoms, and those, whose chains do not have non-contiguous residue numbers~\cite{Hinsen:2013}. We thank K.~Hinsen for sharing with us the results of the checks.} For this set of 212 selected proteins we will construct the Frenet framing, determine the local rotation angles and compute topological indices. We will demonstrate the statistics of the indices for all the studied proteins and consider a couple of examples in more detail.

\begin{figure}[t]
\begin{minipage}{0.45\linewidth}
\includegraphics[width=\linewidth]{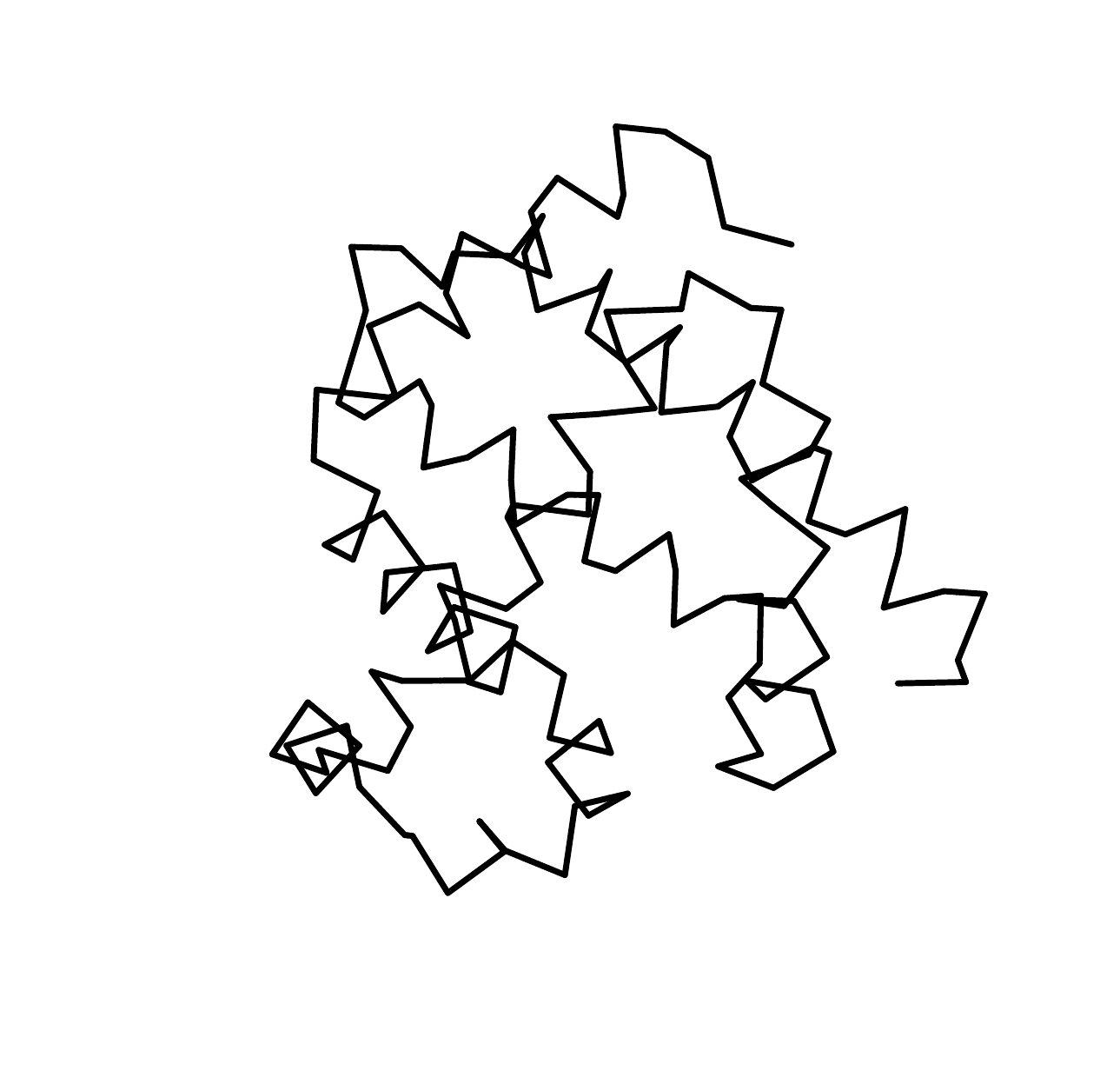}
\end{minipage}
\hfill{
\begin{minipage}{0.45\linewidth}
\includegraphics[width=0.45\linewidth]{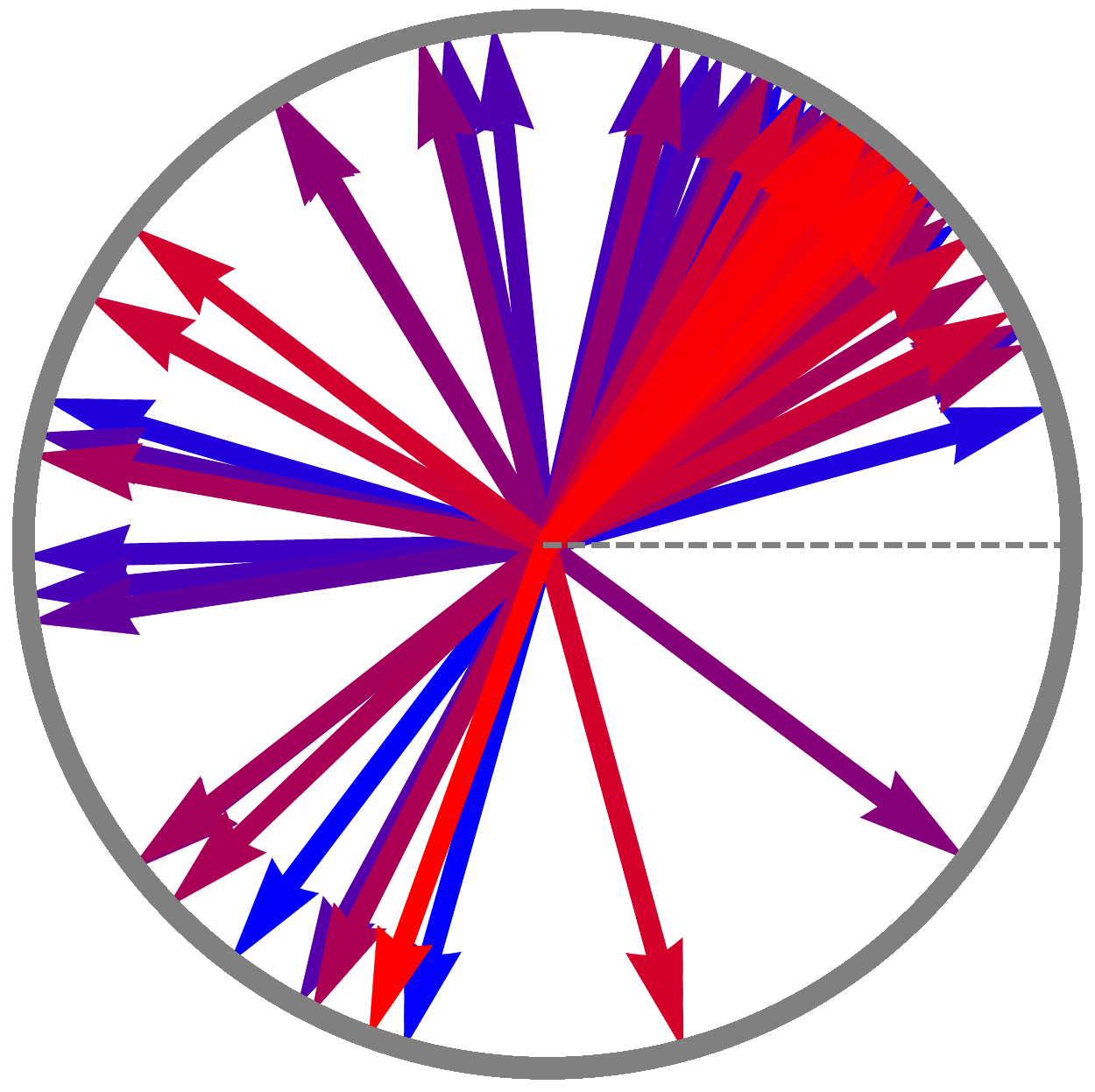}
\hfill{
\includegraphics[width=0.45\linewidth]{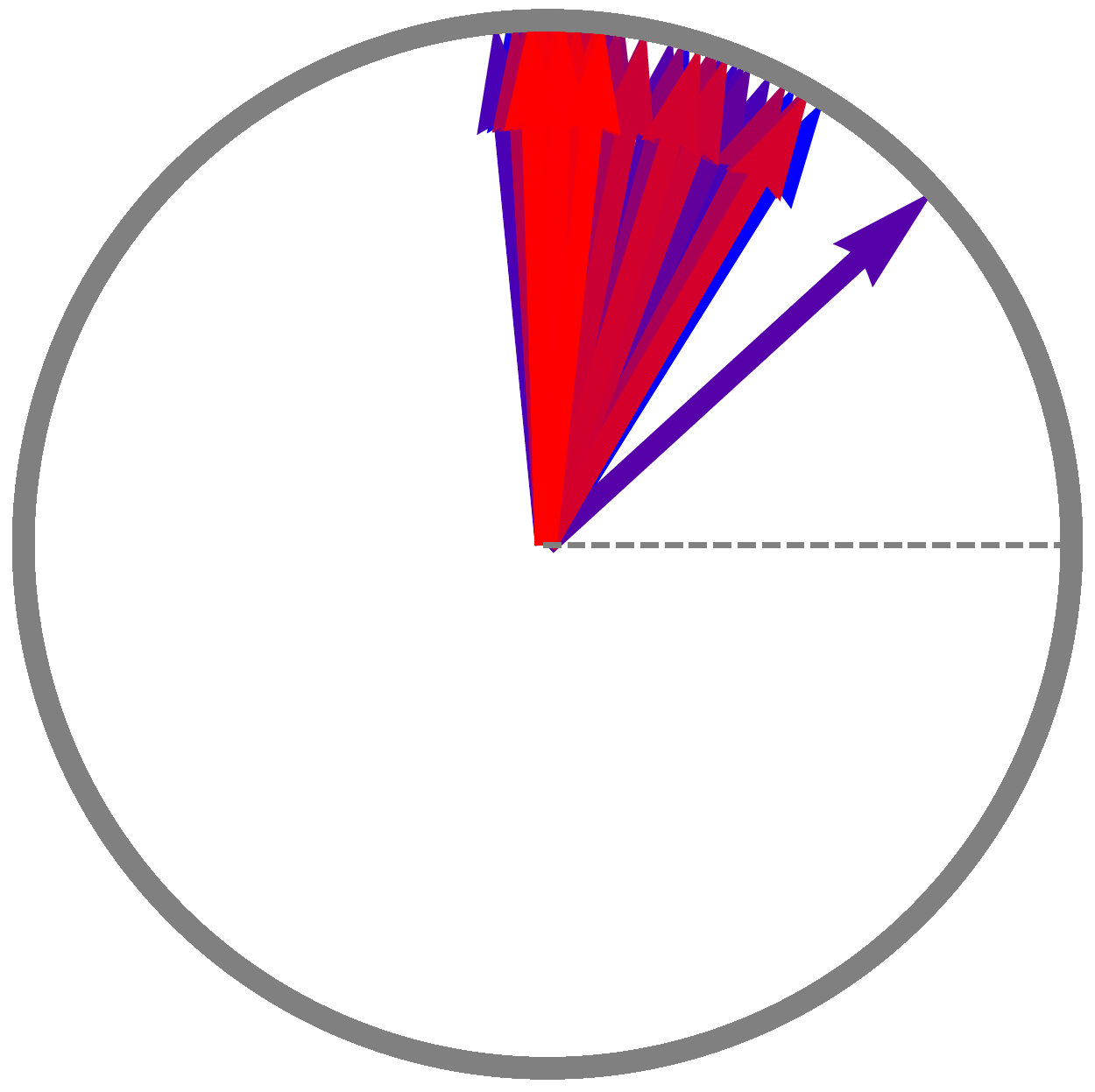}
}
\end{minipage}
}
\caption{ (Left) Chain of the $C_\alpha$ atoms in myoglobin molecule, \emph{1a6m}. (Right) Clocks representing torsion angles $\tau_n$ and curvature angles $\kappa_n$. Color coding shows the order of rotations, with more blue ones occurring earlier than mode red ones.}
\label{fig:1a6m}
\end{figure}

We start with an example of the myoglobin, code \emph{1a6m} in the Protein Data Bank (PDB). Figure~\ref{fig:1a6m} shows the $C_\alpha$ chain of this protein and the set of all curvature and torsion angles. The angles are mapped to the phases of unit vectors on a complex plane. 

In the case of torsion, 124 out of the total 148 plotted vectors on the clocks of figure~\ref{fig:1a6m} point to a direction between zero and $\pi/2$ with the average value of $\bar{\tau}\simeq 50^\circ$. The remaining 24 vectors reflect few possible deviations of the rotation from this standard angle.  Behavior of the curvature angles is even more regular. Most of the rotations are concentrated close to $90^\circ$ with the average value of $\bar{\kappa}\simeq 86^\circ$.

Similar behavior of the curvature and torsion angles can be observed for the remaining proteins. Therefore, we would like to work with the following simple model. We will consider a protein as a set of regular periodic structures connected by ``kinks" -- some irregular connections~\cite{Lundgren:2013}. Regular structures are characterized by a uniform curvature and torsion angles $\bar{\kappa}$ and $\bar{\tau}$, while angles, considerably deviating from those, correspond to the kinks. This correlates with the common secondary structure classification of proteins, with regular structures corresponding to helices, and kinks -- to structural motifs connecting them. We like to think of regular structures as of a ``ground state" of the polygon chain associated with the protein molecule, and of the kinks -- as of  walls separating domains of the ground state.


\subsection{Torsion and curvature indices}

Basing on the simple view of proteins given above, we would like to introduce a topological classification of the irregular parts of a protein, that is a characteristic that will not care that much about the absolute positions of the kinks, but rather about their sequence and their intrinsic features. One such classification can be produced by a number of full rotations that the clock vectors representing the torsion angle (\emph{e.g.} on figure~\ref{fig:1a6m}) make around the center of the circle, as one follows the polygon chain.  

Given the set of torsion angles~$\tau_i$ one computes the following quantity 
\be
\label{tauindex}
\vartheta \ = \ \frac{1}{\pi}\sum\limits_{i=1}^{N-4}\Delta\tau_{i}\,.
\ee
This index, first introduced in~\cite{Lundgren:2013}, computes the sum of planar rotations of $\tau_{n}$ vectors on figure~\ref{fig:1a6m}. The rotations are taken in the direction of the smaller angle between $\tau_n$ and $\tau_{n+1}$. Using the coordinate presentation of vectors, $\vec{\tau}_i=(\cos\tau_i,\sin\tau_i,0)$, the change of angle is defined through 
\be
\label{DeltaTau}
\Delta\tau_i= \sign\left(\hat{\bf z}\cdot(\vec{\tau}_i\times\vec{\tau}_{i+1})\right)\arccos\left(\vec{\tau}_i\cdot\vec{\tau}_{i+1}\right)\,,
\ee
where $\hat{\bf z}=(0,0,1)$ denotes a unit vector perpendicular to the plane. 

\begin{figure}[t]
\begin{minipage}{0.45\linewidth}
\includegraphics[width=\linewidth]{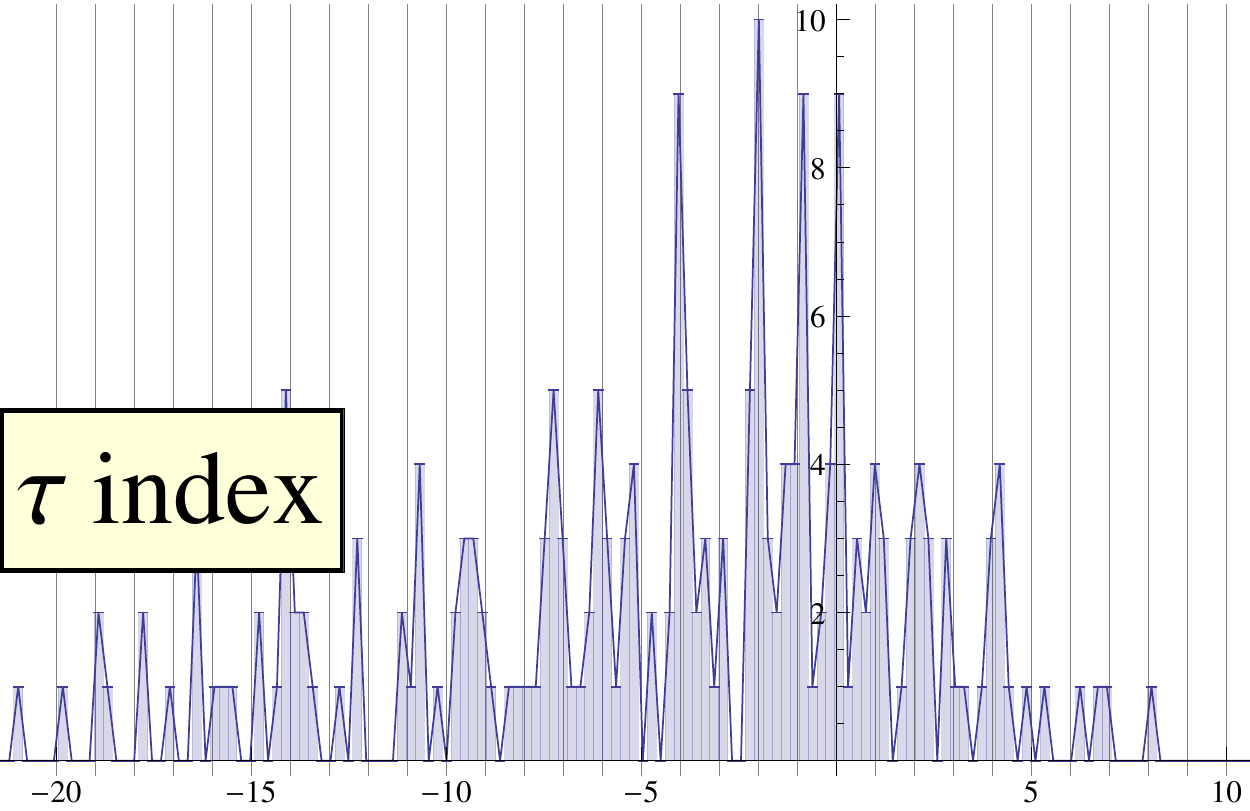}
\end{minipage}
\hfill{
\begin{minipage}{0.45\linewidth}
\includegraphics[width=\linewidth]{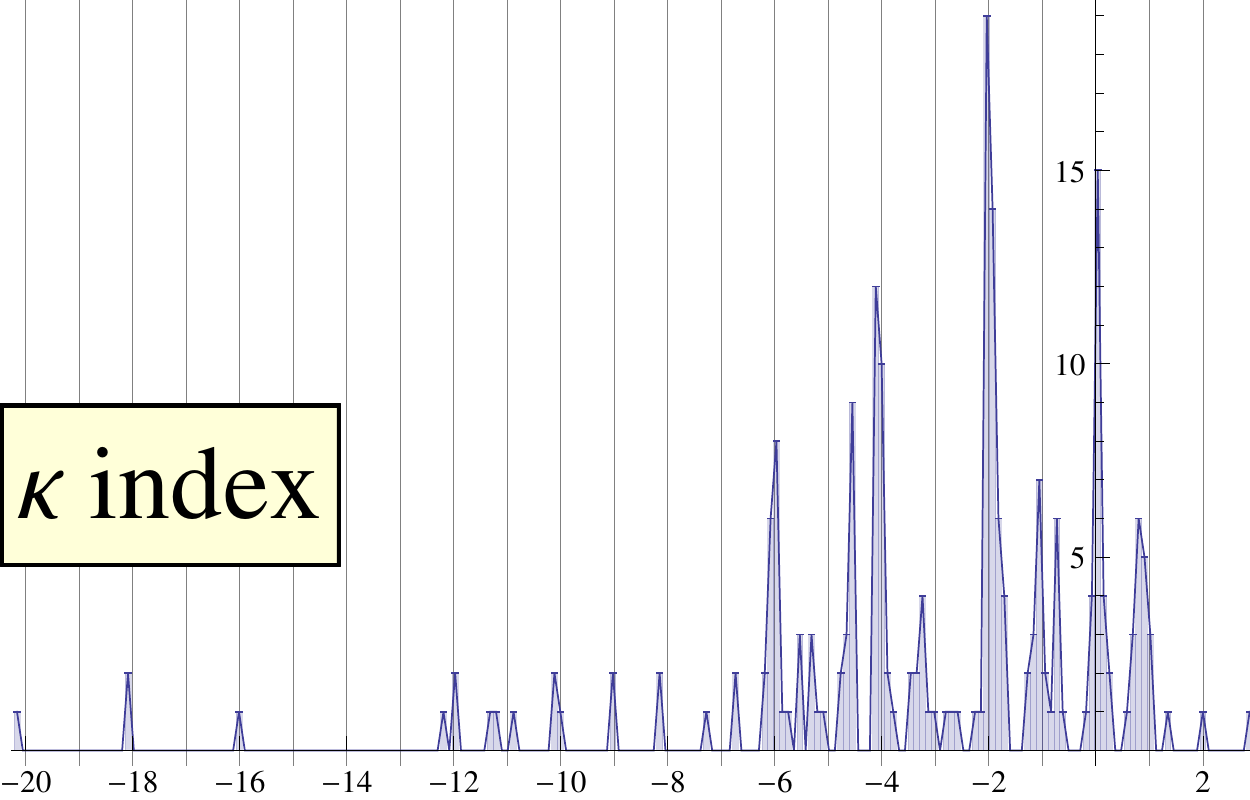}
\end{minipage}
}
\caption{(Left) Distribution of the torsion index $\vartheta$ showing the winding of the $\vec{\tau}_i$ trajectory. (Right) Distribution of the curvature index $\varpi$ showing the winding of the $\vec{\kappa}_i$ trajectory.}
\label{fig:indices}
\end{figure}

The analysis of 212 selected proteins is summarized by the histogram on figure~\ref{fig:indices} (left), which shows the distribution of the quantity $\vartheta$. As can be seen from the figure, it has a high propensity towards integer values. This quantization was first discussed in~\cite{Lundgren:2013}, where $\vartheta$ was called \emph{folding index}. 

Note that instead of summing $\Delta\tau_i$ we could instead sum the angles $\tau_i$ themselves. Such an index would be an analog of the integral of the torsion in the continuous case. The latter is the twist $tw$, which appears in the Calugareanu's theorem~(\ref{Calugareanu}) and is not in general an integer number. Index $\vartheta$ computed here, is a discrete version of the integral of the derivative of the torsion, which has loci on the irregular pieces of the chain. It is integer in units of $\pi$.

\begin{figure}[t]
\begin{minipage}{0.45\linewidth}
\includegraphics[width=\linewidth,clip=true,trim=0pt 50pt 0pt 50pt]{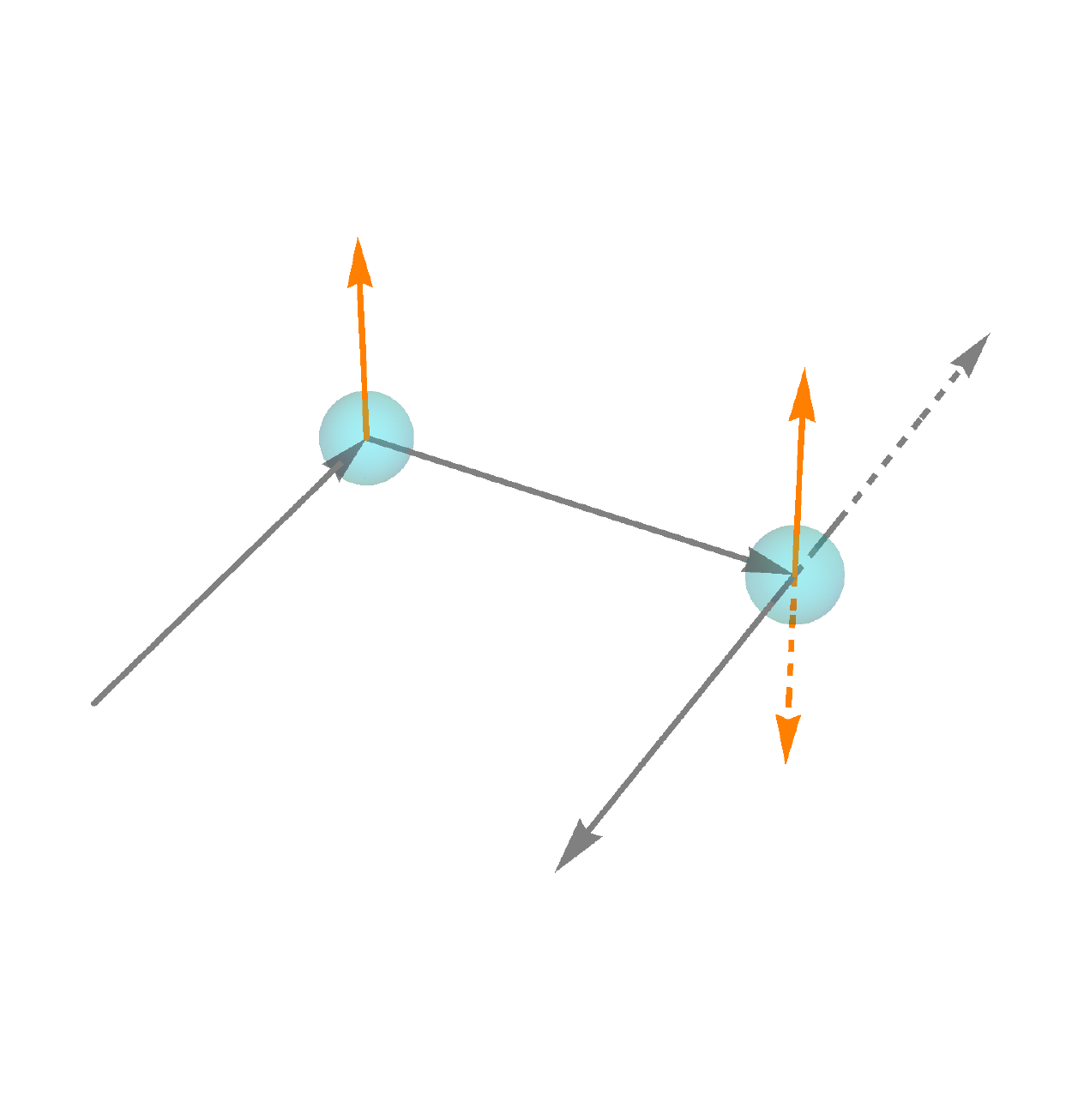}
\end{minipage}
\hfill{
\begin{minipage}{0.45\linewidth}
\centering
\includegraphics[width=0.6\linewidth]{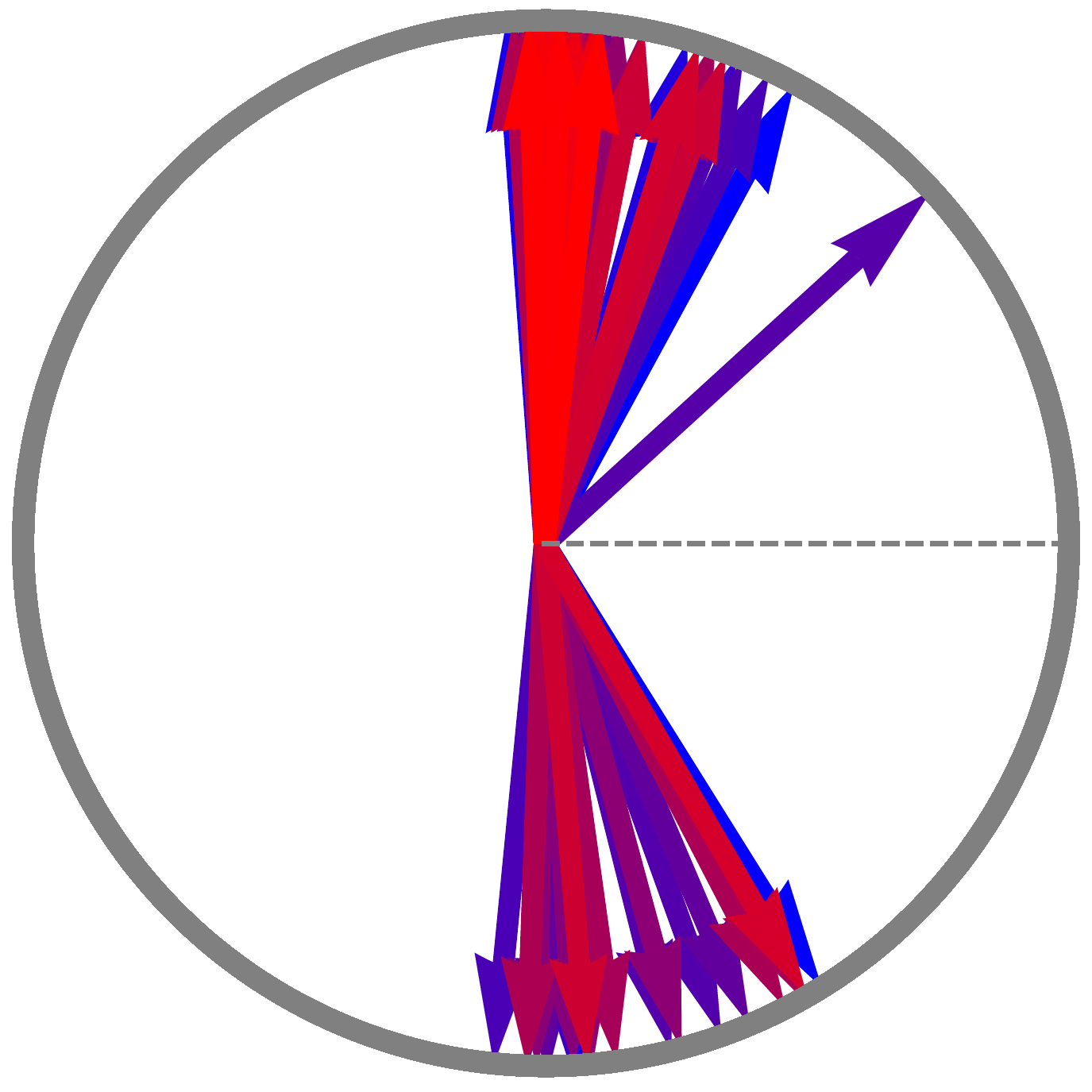}
\end{minipage}
}
\caption{(Left) chain without and with inflection (dashed vectors). (Right) 
Distribution of $\kappa_n$ curvature angles in the myoglobin taken with a sign.}
\label{fig:indcor}
\end{figure}

Do the curvature angles carry any similar topological information? A calculation using definition~(\ref{tauindex}), but with $\Delta\kappa_i$ instead of $\Delta\tau_i$ would produce a trivial answer. Few reasons for that is that $\kappa_n$ are restricted to take values in $[0,\pi]$ and all fluctuate close to the same value $\pi/2$. In particular, $\kappa_n$ do not distinguish two situations shown on figure~\ref{fig:indcor} (left), where the chain has two alternative opposite directions. The alternative, shown by dashed vectors, is an inflection, characterized by a flip of the orientation of the binormal vector (orange). In order for inflections to be taken into account by $\kappa_n$ we decorate the angles with an additional sign, which is determined as follows.

For the first frame we assign $\kappa_1$ to be positive. At any position $n+1$ the sign is defined with respect to the relative orientation of the consecutive $\bb$ vectors: $\sign(\kappa_{n+1})=\sign(\kappa_{n})\sign(\bb_n\cdot\bb_{n+1})$. That is the sign at $n+1$ remains the same as the sign at $n$, if the angle between vector $\bb_n$ and $\bb_{n+1}$ is less than $90^\circ$ and vice versa. Note that modified $\kappa_n$ is no longer as local as the original one: it requires at least two consecutive binormal vectors. There is a corresponding global $Z_2$ symmetry that distinguishes two choices of the sign. This symmetry is spontaneously broken forming domains of different signs of curvature angles. In the myoglobin molecule, the distribution of the curvature angles calculated with the sign is shown on figure~\ref{fig:indcor} (right). In this diagram 21 out of 148 curvature angles have negative sign.

Given the set of oriented $\kappa_i$ one can now calculate  index
\be
\varpi \ = \ \frac{1}{\pi}\sum\limits_{i=1}^{N-4}\Delta\kappa_{i}\,,
\ee
with $\Delta\kappa_i$ defined in a similar way to $\Delta\tau_i$, \emph{cf.} Eq.~(\ref{DeltaTau}). Analyzing 212 proteins we found the distribution of index $\varpi$ shown on the right panel of figure~\ref{fig:indices}. Again, the index appears to be ``quantized" in units of $\pi$.

Above we have compared index $\vartheta$ with twist $tw$. One can observe that the second index is similar to writhe.  Indeed,  it is clear that for a straight line, the tangent vector is parallel to itself, the normal vectors rotate in a perpendicular plane and a full $2\pi$ rotation adds or removes a unit of twist. Similarly, for a closed planar curve the binormal vector is always parallel to itself and a full $2\pi$ rotation gives a shift of the writhe by one unit. Similarly to $\vartheta$, index $\varpi$ computes the relative contribution of large inhomogeneous rotations of the $\kappa$ vector around the origin.

\subsection{Evolution of indices and kink structures}

\begin{figure}[t]
\begin{minipage}{0.45\linewidth}
\includegraphics[width=\linewidth]{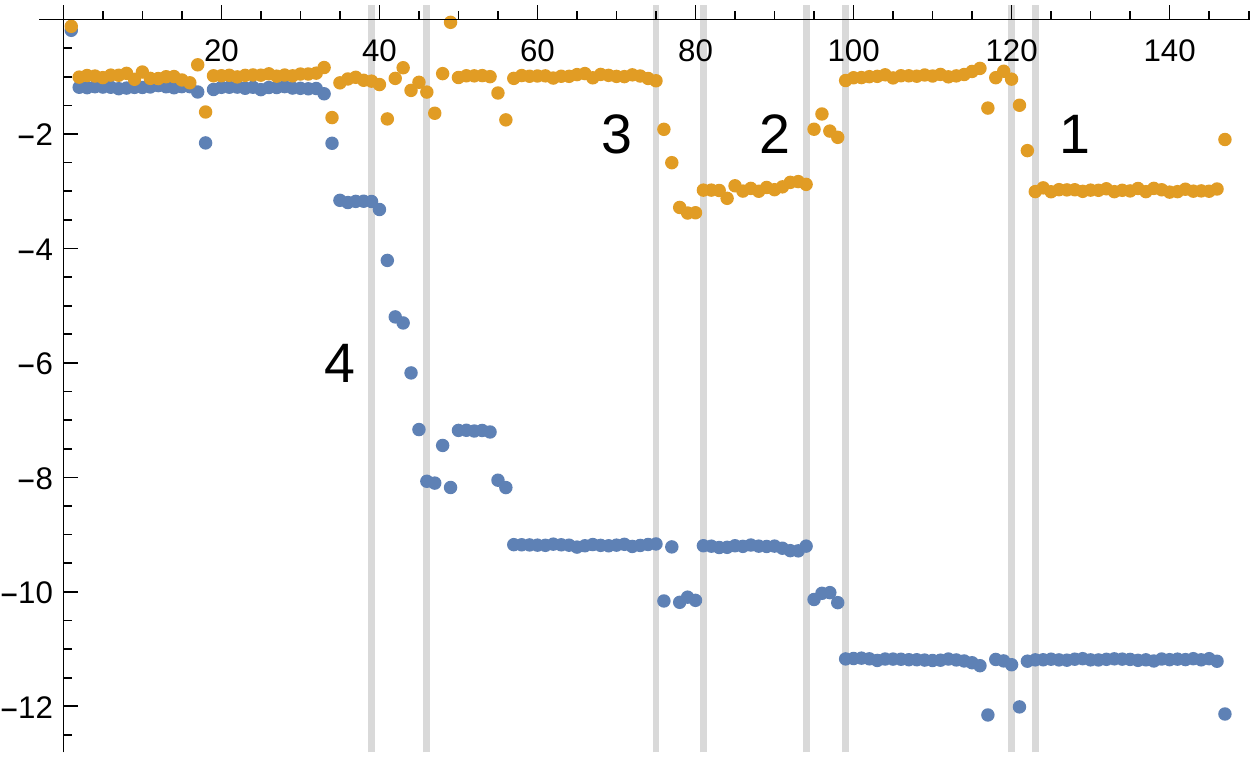}
\end{minipage}
\hfill{
\begin{minipage}{0.45\linewidth}
1 \includegraphics[width=0.45\linewidth]{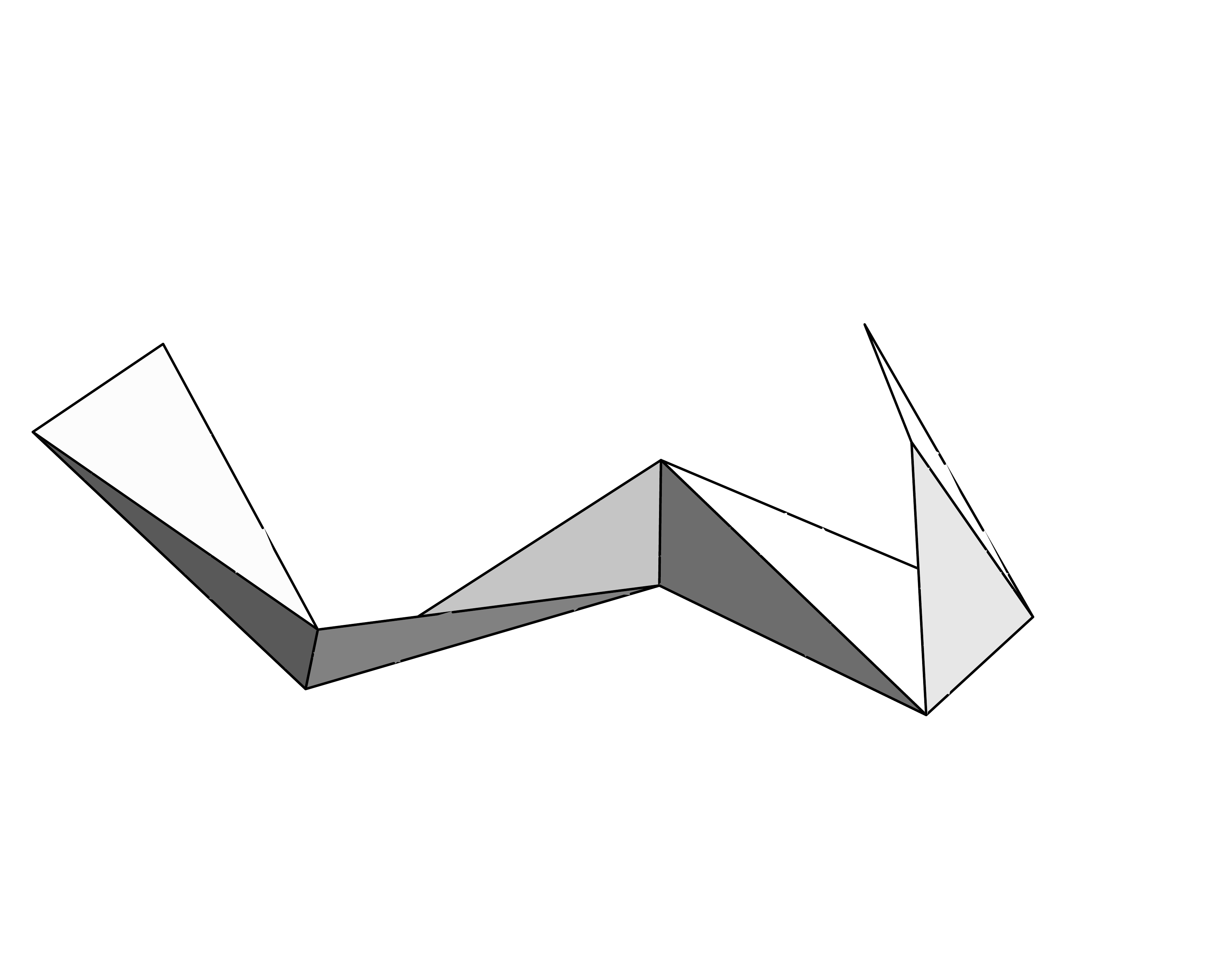}
\hfill{
2\includegraphics[width=0.45\linewidth]{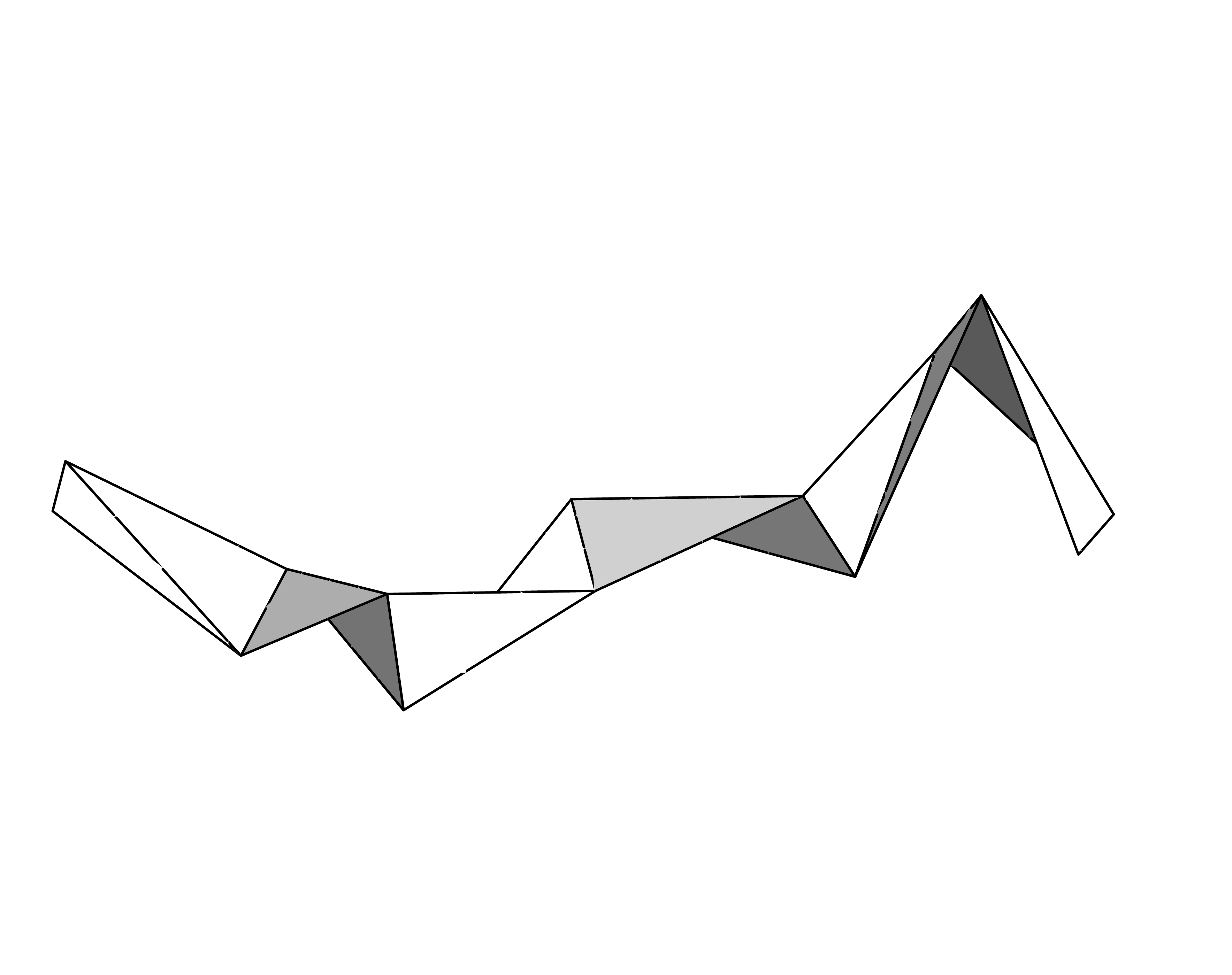} \\
3\includegraphics[width=0.45\linewidth]{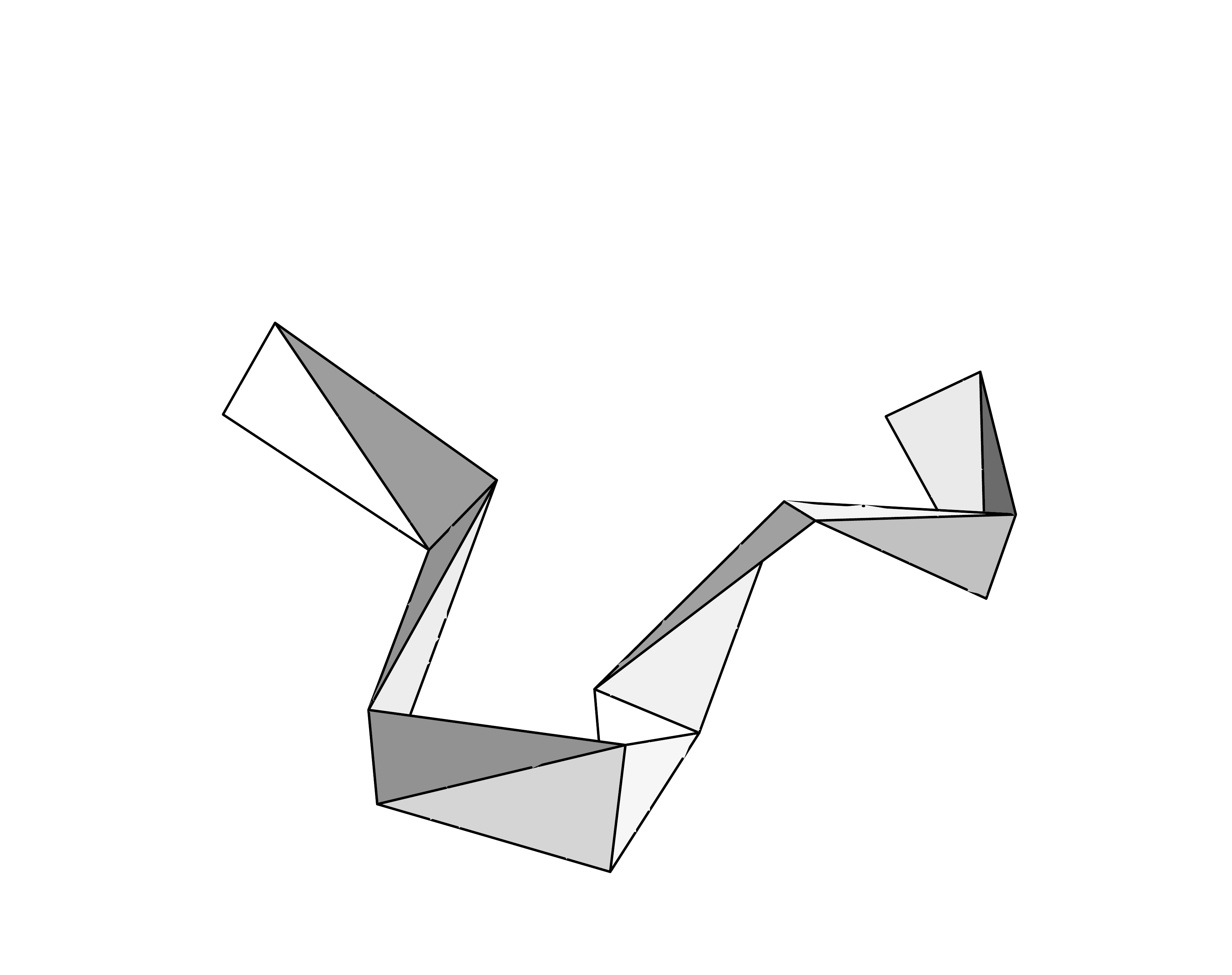}
\hfill{
4\includegraphics[width=0.45\linewidth]{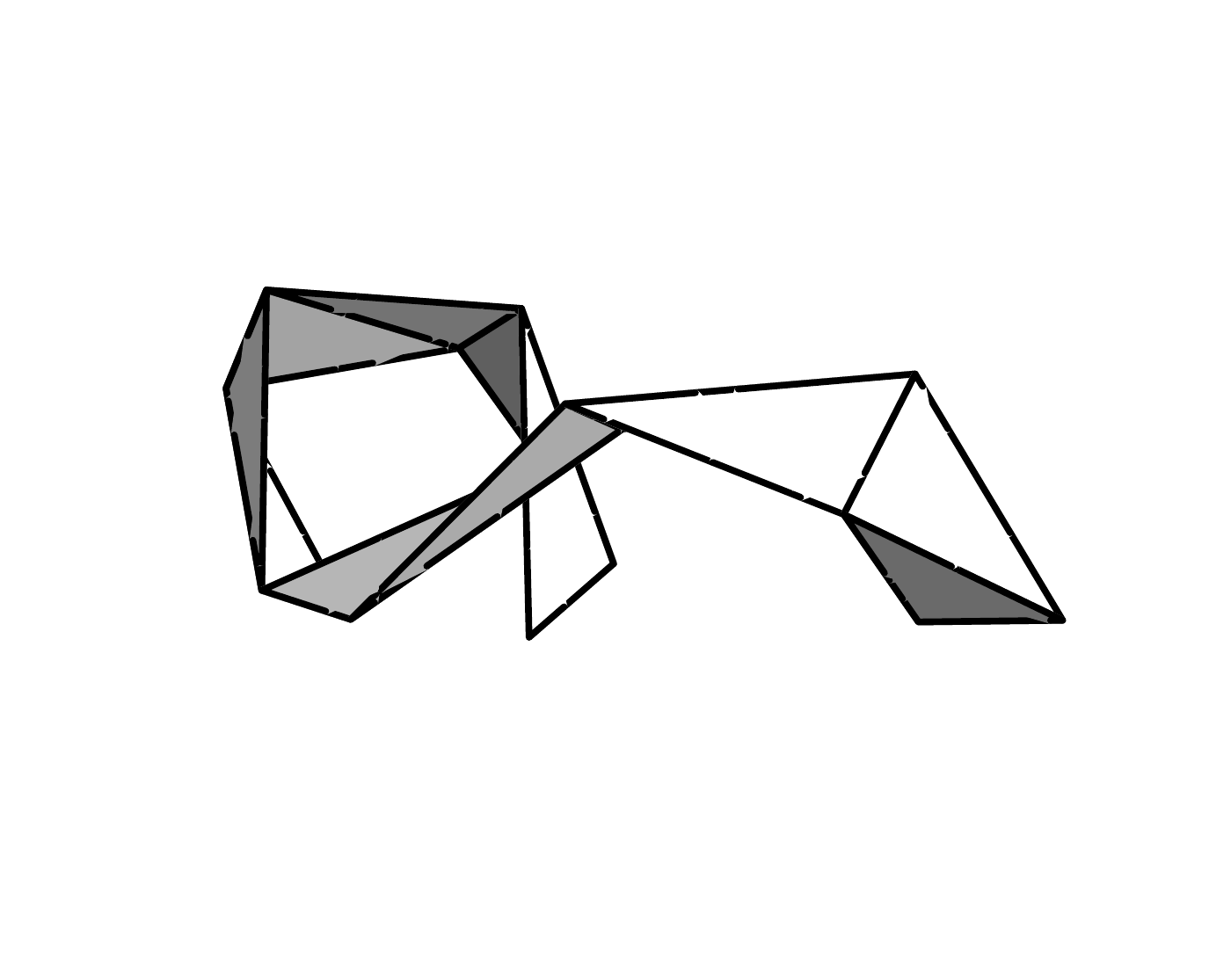}
}
}
\end{minipage}
}
\caption{Accumulation of $\vartheta$ (yellow) and $\varpi$ (blue) indices along the chain of myoglobin ({\sl 1a6m}). Horizontal axis labels the amino acid sequence. Insets show the shapes of framed polygons at the labeled positions.}
\label{fig:1a6mindex}
\end{figure}

We can look a little closer at the protein data to see how indices $\vartheta$ and $\varpi$ are built. We consider two characteristic examples. The first one, again, is the myoglobin (\emph{1a6m}), which is a helical protein. The regularity of its structure can be observed from the plot of the index accumulation on figure~\ref{fig:1a6mindex}. The plot shows the regions of constant index, which correspond to helical structures, and irregular connections through which the index jumps.

\begin{figure}[t]
\begin{minipage}{0.45\linewidth}
\centering
\includegraphics[width=\linewidth]{./tw1a6m.pdf} \\

\vspace{-1.cm}
\includegraphics[width=0.3\linewidth]{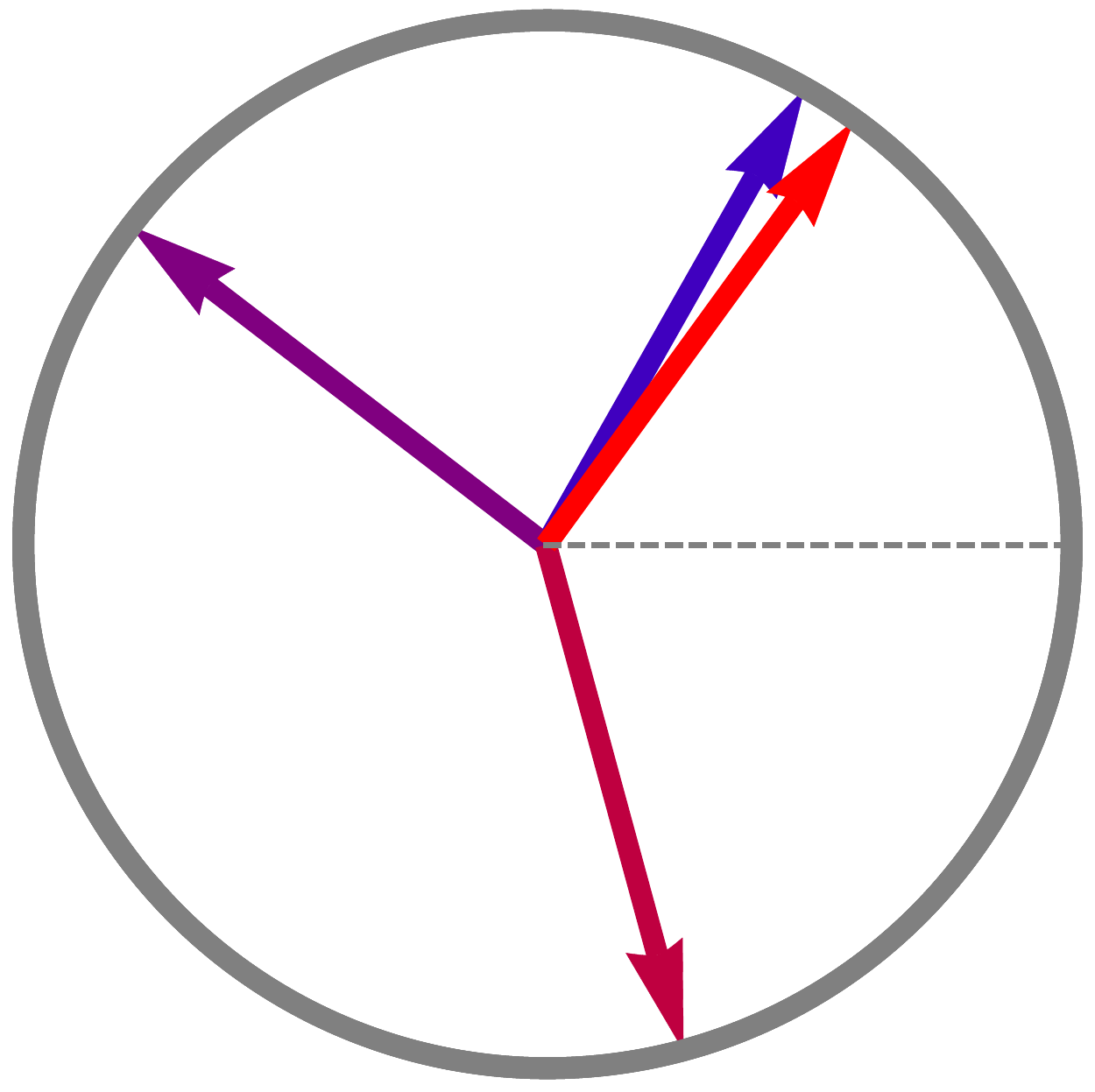}
\qquad
\includegraphics[width=0.3\linewidth]{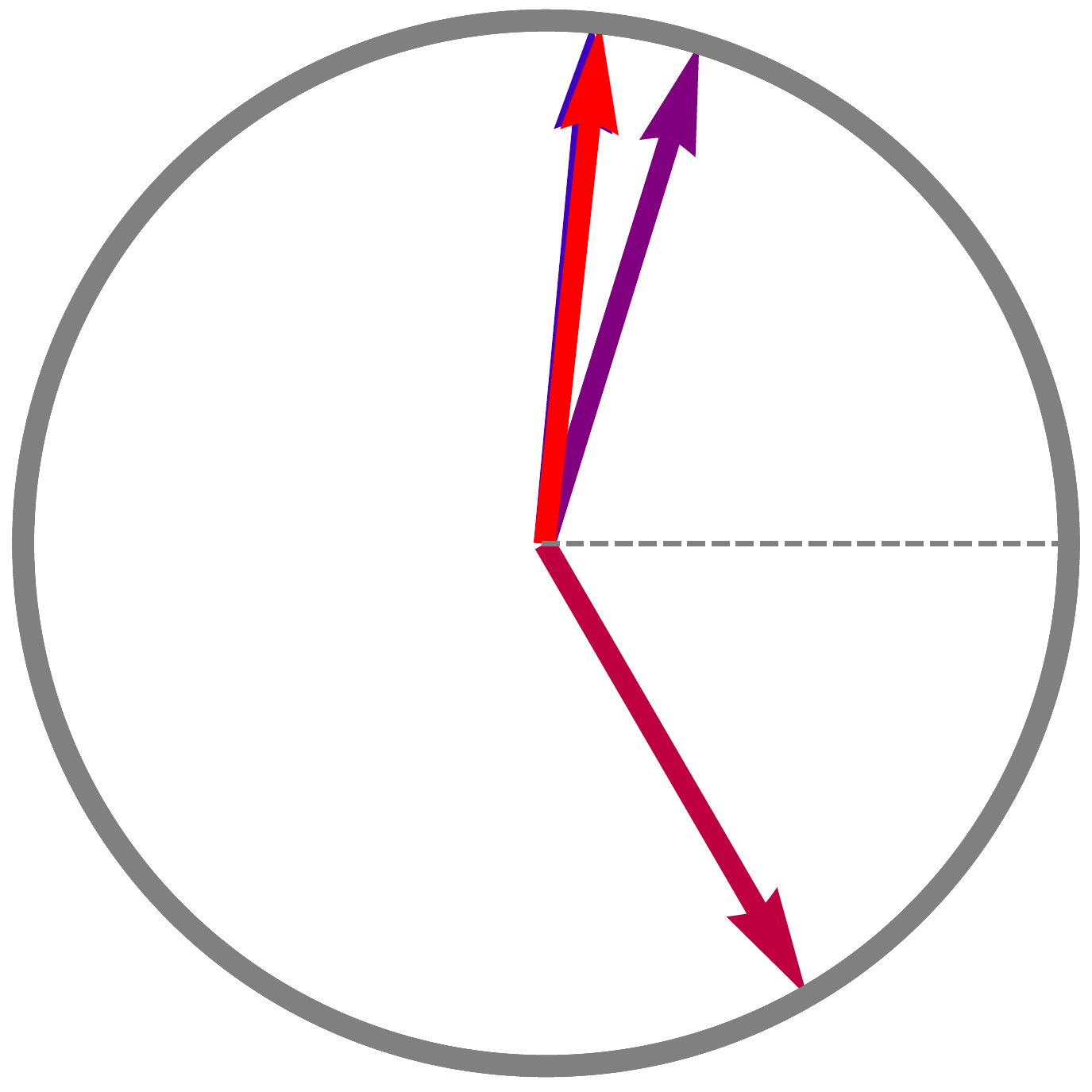}
\end{minipage}
\hfill{
\begin{minipage}{0.45\linewidth}
\centering
\includegraphics[width=\linewidth]{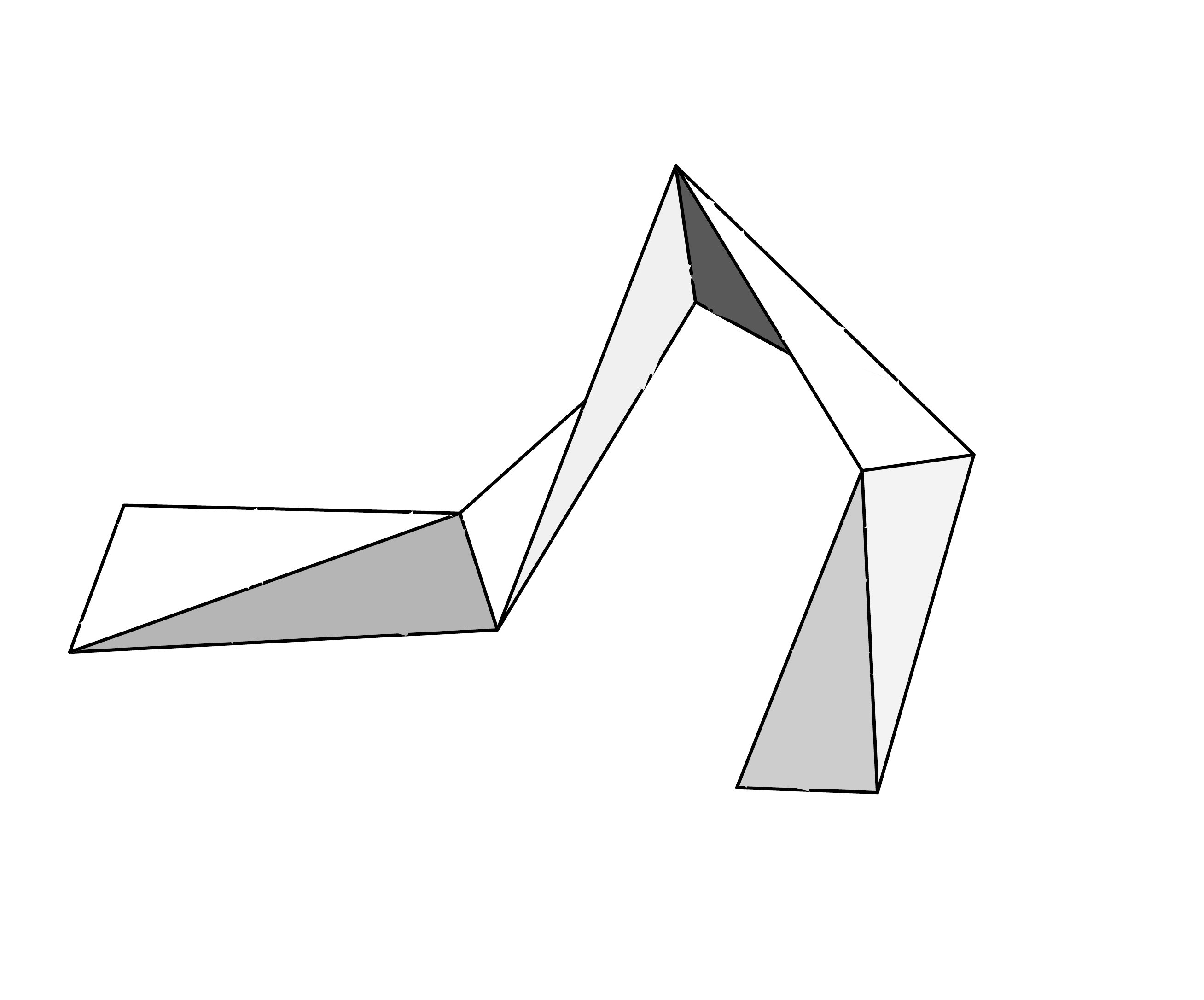} \\

\vspace{-1.cm}
\includegraphics[width=0.3\linewidth]{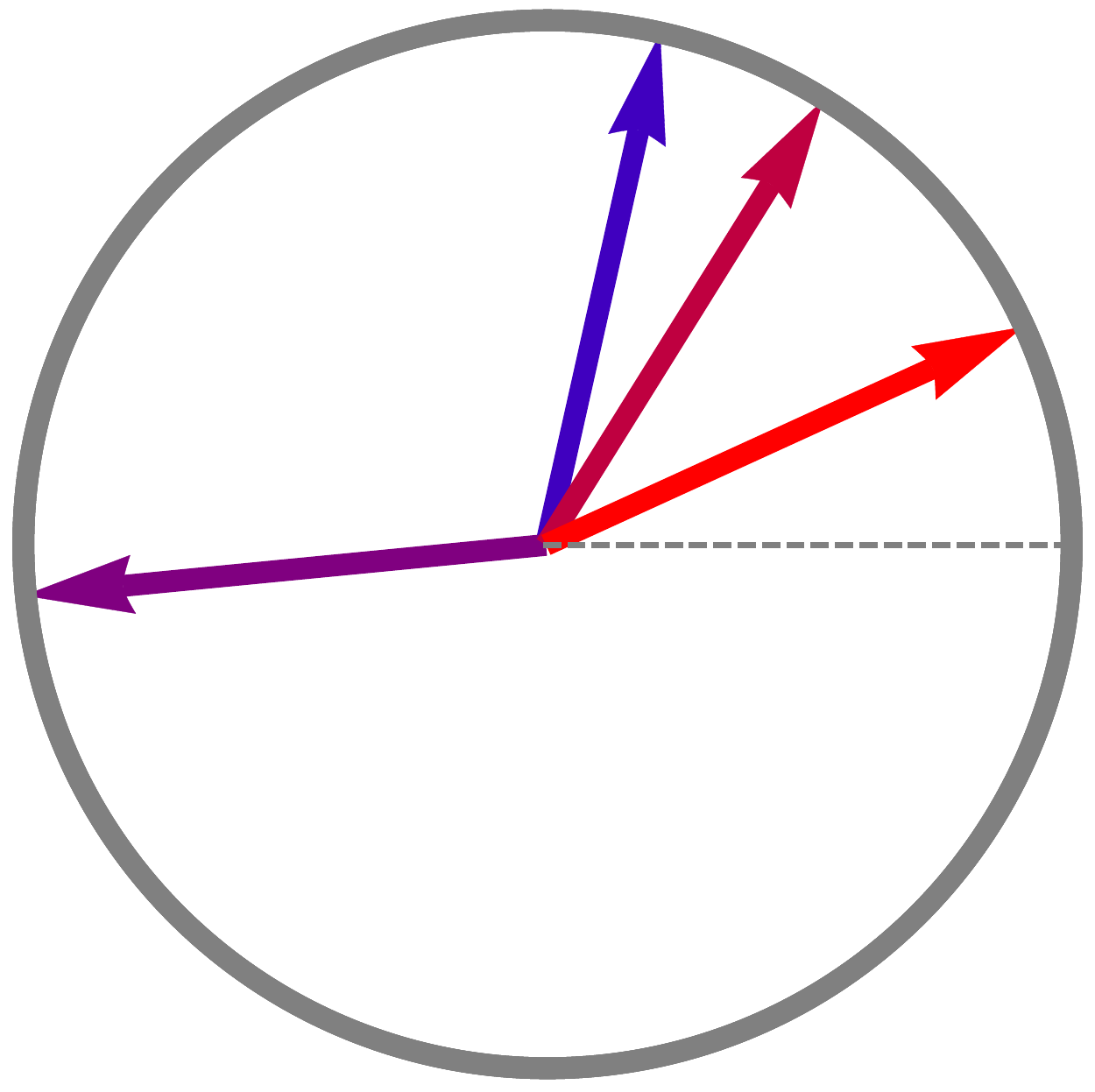}
\qquad
\includegraphics[width=0.3\linewidth]{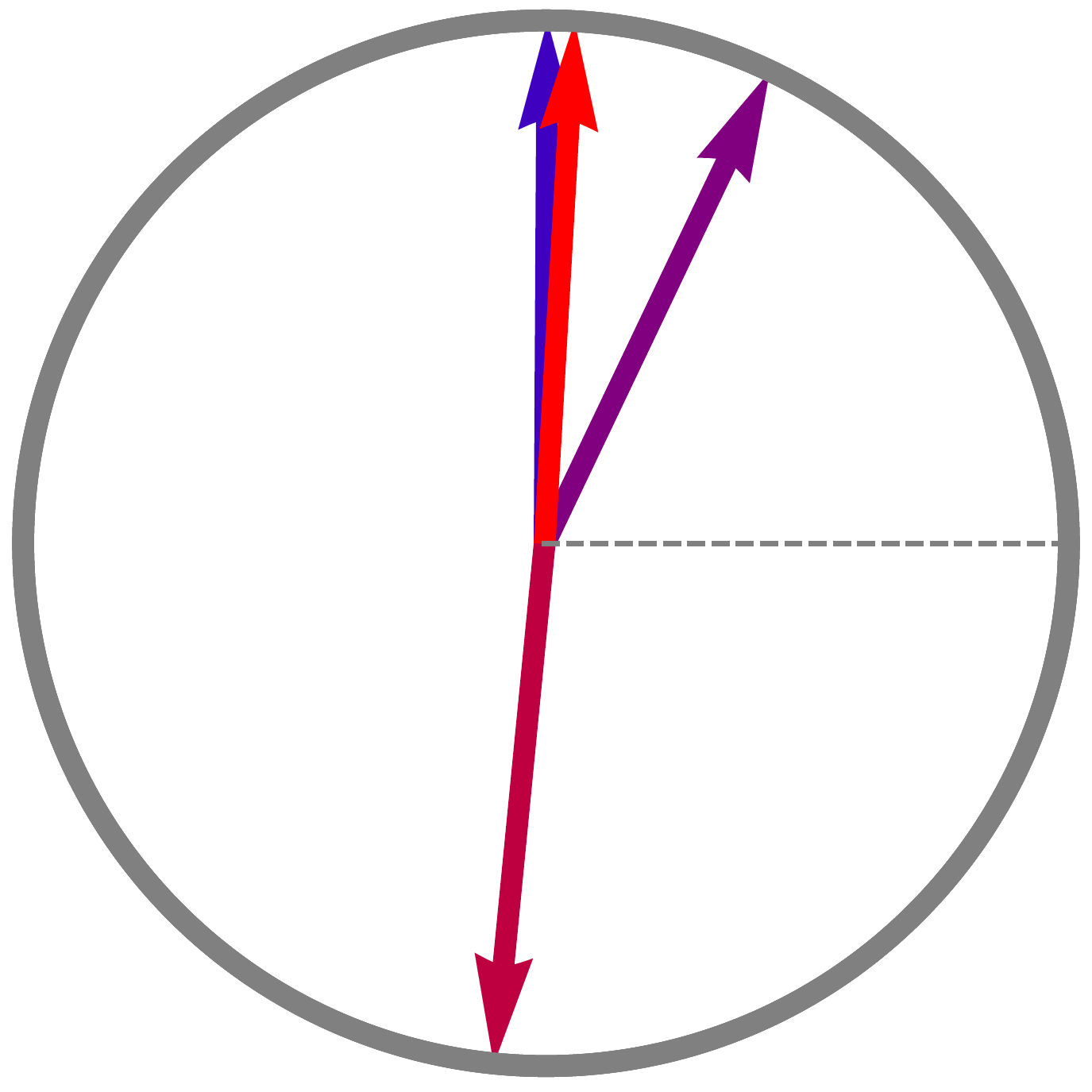}
\end{minipage}
}
\caption{Jump in $\vartheta$ (left) and jump in $\varpi$ (right). In each case clock diagrams show the sequence of $\tau_n$ (left) and $\kappa_n$ (right).}
\label{fig:difference}
\end{figure}

One might be interested in how the loci of index accumulation may look like. This is shown on the insets on figure~\ref{fig:1a6mindex}, which visualize framed kinks at the selected locations along the chain. More specifically, from the left panel of figure~\ref{fig:difference} one can see how the unit of index $\vartheta$ is obtained, while the right panel of the same figure shows the locus of the $\varpi$ index.

The second characteristic example is shown on figure~\ref{fig:1i1w} (left) with the xylanase protein~(\emph{1i1w}). This is a longer protein with a richer structure. In contrast to \emph{1a6m}, there is a small variation of $\varpi$ index, while $\vartheta$ index exhibits cascades of significant changes.

One can notice already from the two examples considered here that the two indices are not quite correlated. This can also be observed in the plot of $(\vartheta,\varpi)$ pairs for many proteins on figure~\ref{fig:1i1w} (right). In the next section we will discuss a more precise connection of the indices with the self-linking number discussed before.


\section{Self-linking number}
\label{sec:main}

\subsection{Twist}

Consider the following topological model of proteins. The protein backbone can be mapped to a chain~(\ref{DiscretePexp}) of elementary rotations~(\ref{elemrot}). At regular positions, corresponding to helices, one inserts a rotation matrix $S_{n+1,n}$ with uniform rotation angles $\bar\kappa$ and $\bar\tau$, as in the previous section. For a small number of positions non-uniform rotations are inserted. In a simple model, we will assume that the non-uniform rotations introduce an additional shift of the $\tau$ angle by $\pm\pi$. In other words, we will assume the following form of the rotation matrix at every position
\be
\label{spinchain}
\hat{S}_{n+1,n} \ = \ 
{\rm e}^{ i\theta_n\pi \hat{t}_{n+1}/2}\cdot S_{n+1,n} 
\,, \qquad \pi\theta_n \ = \ \tau_n - \bar\tau\,.
\ee   
In the present model $\theta_n$ can take values 0, 1 and $-1$, so that at every position there is either a uniform rotation, or a rotation with an additional shift.

It follows from relation~(\ref{Frenet2}) between frames and rotation matrices that
\be
\label{commutator}
S_{n+1,n}\cdot\hat{e}_a(n) \ = \ \hat{e}_a({n+1})\cdot S_{n+1,n}\,.
\ee 
In other words,  rotation matrix $S_{n+1,n}$ transports the frame at position~$n$ to the frame at position $n+1$.  We can view the extra rotation around $\tb_{n+1}$ in equation~(\ref{spinchain}) as a component of a frame following
\be
\label{pirot}
\exp\left(\pm i\,\frac{\pi}{2}\,\hat{t}_n\right) \ = \ {{\rm e}^{\pm i \pi/2}}\,\hat{t}_n\,,  \qquad \exp\left(\pm i\,\frac{\pi}{2}\,\hat{b}_n\right) \ = \ {\rm e}^{\pm i \pi/2}\,\hat{b}_n\,.
\ee
Hence we can apply equation~(\ref{commutator}) replacing $\hat{e}$ by $\hat{t}$: matrices $S_{n+1,n}$ will then transport $\tb_n$ at position $n$ to $\tb_{n-1}$ at position $n-1$ and vice versa. 

In the string of discrete rotations~(\ref{DiscretePexp}), one can commute the large $\pi$ rotations through the chain of regular rotations and collect them in the beginning of the chain, arriving at a string 
\be
\label{LkFactor}
S(N) \ = \  {\rm e}^{i \frac{\pi}{2}\theta}\cdot S_{N-2,N-3}\cdots S_{3,2}\cdot S_{2,1}\cdot \hat{t}_1^{\theta}\,,
\ee
where 
\be
\theta \ = \ \theta_2+ \ldots + \theta_{N-2}\,.
\ee  
In equation~(\ref{LkFactor}) the full rotation matrix splits into three factors. The last factor is an overall $\theta\pi$ rotation of the first frame $\hat{e}_a(1)$ around the tangent vector. It comes from commuting all the extra rotation matrices through the uniform rotations $S_{n+1,n}$.  Since $\hat{t}_1^2=1$, it is enough to consider $\theta$ modulo two in this factor. 

The rotation of $\hat{e}_a(1)$ comes with a phase -- the first factor in  equation~(\ref{LkFactor}). We understand the phase $\theta$ as a (half of the) discrete self-linking number associated with the large $\pi$ rotations of the Frenet frame. It is an integer number and being associated with rotations around a tangent vector it is a discrete contribution to ``twist'' $tw$. While a $\pi$ rotation of a frame as vectors is ambiguous, the spinor representation distinguishes the direction of such a rotation.

Finally, there is a factor corresponding to a chain of uniform rotations that computes the self-linking number of a helix,
\be
\label{Lk0}
 S_{N-2,N-3}\cdots S_{3,2}\cdot S_{2,1} \ = \ S_{N-2,1} \ = \  {\rm e}^{\pi i Lk_0\hat{t}'} \,,
\ee
where $\tb'$ is the ``instantaneous" axis of rotation of the first frame to the last frame.

Let us consider the case of a closed chain. For example, we can assume identifying initial and final points $N\sim 1$. In this case there will be $N-1$ tangent vectors and $N-1$ frames. An additional factor~$\hat{S}_{1,N-1}\hat{S}_{N-1,N-2}$ will be added to the chain of rotations~(\ref{DiscretePexp}), for example, by left multiplication. Periodic boundary conditions will then require that the total product in equation~(\ref{LkFactor}) combines into a trivial rotation, that is a plus or minus identity in the spinor representation. The two possibilities are that either rotations around $\tb_1$ and around $\tb'$ are both trivial, producing phases ${\rm e}^{i\pi Lk_0}$ and ${\rm e}^{i\pi/2\theta}$, for even $\theta$, or they are both $\pi$ rotations around $\tb_1$ with $\theta$ being odd and $Lk_0$ half-integer. In either case 
\be
S_{\rm closed} \ = \ \exp(i\pi Lk) \ =  \ {\rm e}^{i \frac{\pi}{2}(\theta+2Lk_0)}\,.
\ee
In other words, we find that the full self-linking number splits in the sum of the regular self-linking number counting the uniform (small) rotations and the large $\pi$ jumps:
\be
\label{Lk}
Lk \ = \ Lk_0 +\frac{\theta}{2}\,.
\ee 
$Lk_0$ is a half-integer if the number of $\pi$ jumps is odd. In most general open chain case $Lk_0$ is a real number, while $\theta$ is an integer. An appropriate generalization of equation~(\ref{Lk}) would be
\be
Lk \ = \ \left[Lk_0\right]+ \left\{\begin{array}{ll}
\frac{\theta}{2}\,, & \text{$\theta$ even}\\
\frac{\theta-1}{2}\,, & \text{$\theta$ odd}
\end{array}
\right.\,,
\ee
where $[Lk_0]$ stands for the integer part of $Lk_0$.

We should yet connect the index $\theta$ discussed here to the index $\vartheta$ computed in section~\ref{sec:indices}. We noted before that the sum of torsion angles, or equivalently, the integral of torsion, is not a topological invariant -- in particular, it is not the full self-linking number. From the point of view of our discussion this happens, because by non-commutativity of three-dimensional rotations, we cannot simply add torsion angles to obtain $Lk$. However, this can be done with the discrete $\pi$ rotations, which ``commute'' with the rotation matrices. Consequently, $\theta$ is counting the sum of the individual jumps accounting for their direction.

Index $\vartheta$ is closely related to $\theta$, but is not precisely the same. It computes local difference between $\tau_n$ and in continuous case would correspond to integrating a derivative of $\tau$. A key subtlety is that torsion angles are defined modulo $2\pi$, to which $\theta$ is less sensitive. To explain this point we first notice that there are two kind of structures in the behavior of $\vartheta(n)$ on figure~\ref{fig:difference} (left): the peaks and the steps. Both of these structures indicate the same jump $\theta_n$ as in the indicated on the following diagram corresponding to a positive jump
\be
\label{interp}
\begin{array}{ccccc}
\text{peak $\vartheta(n)$} & \begin{array}{c}\includegraphics[scale=0.1]{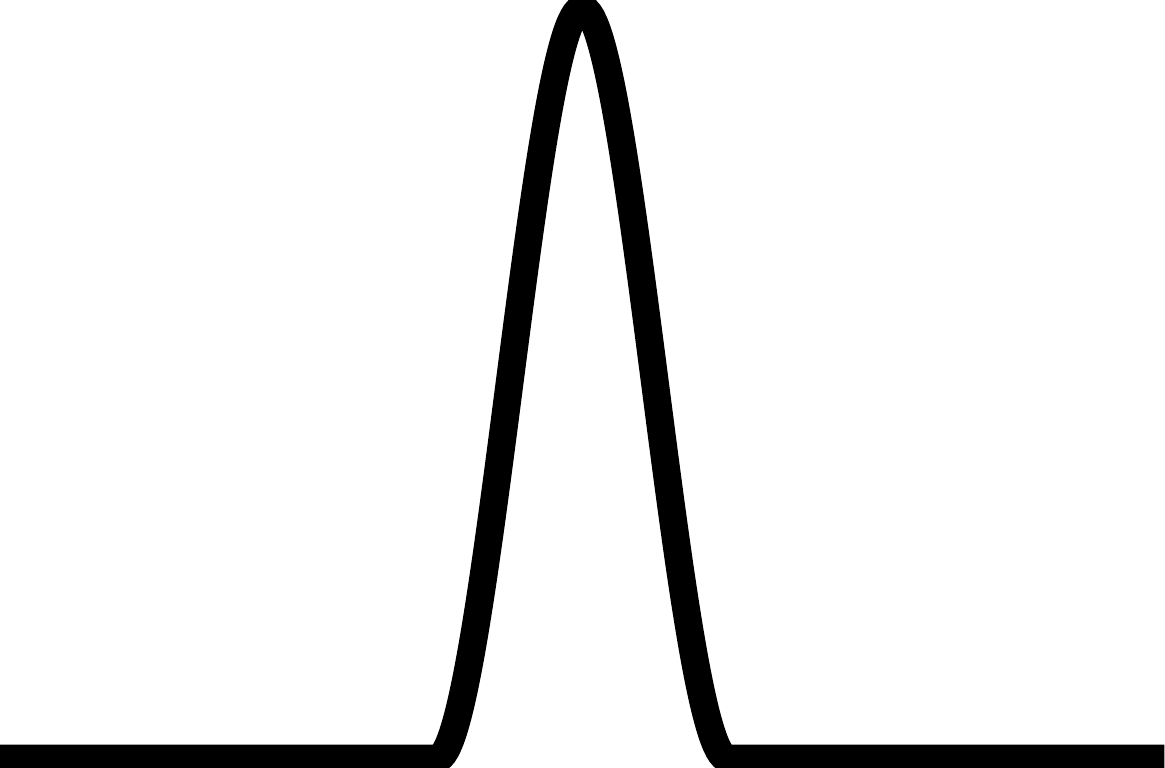}\end{array} & \rightarrow & \text{jump $\theta(n)$} & \begin{array}{c}\includegraphics[scale=0.1]{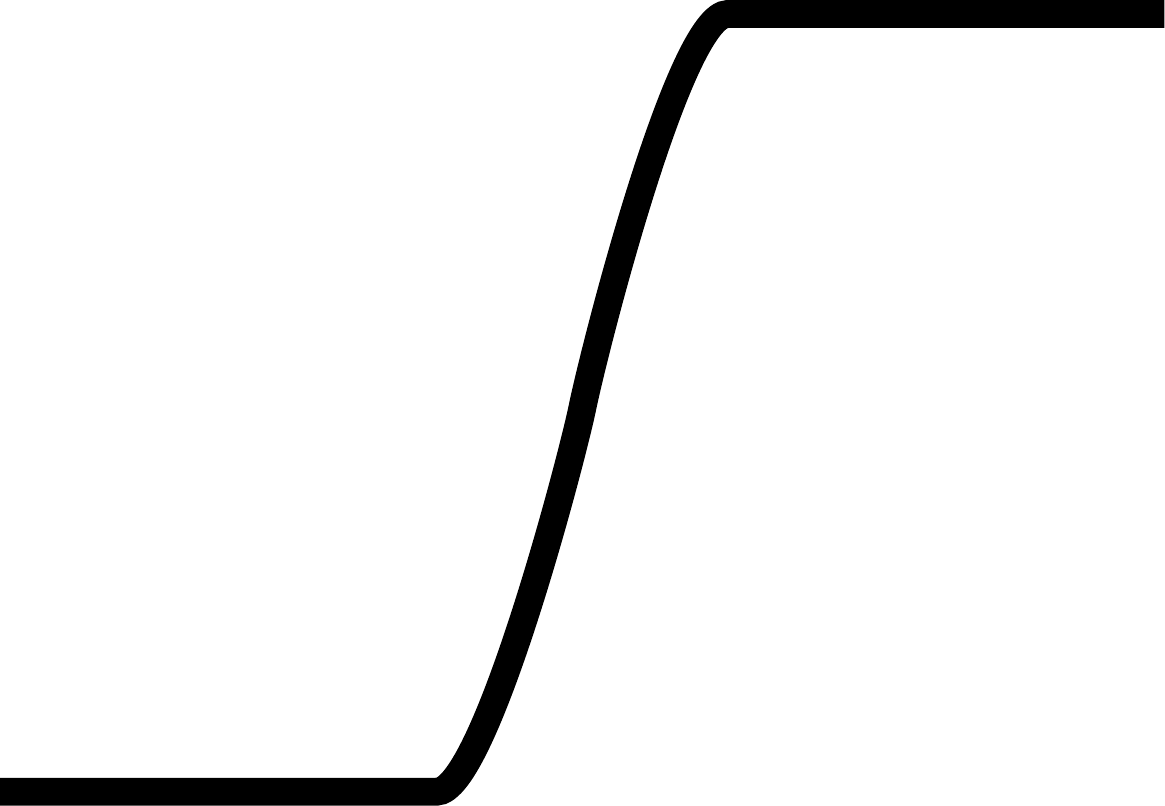}\end{array}\\
\text{step $\vartheta(n)$} & \begin{array}{c}\includegraphics[scale=0.1]{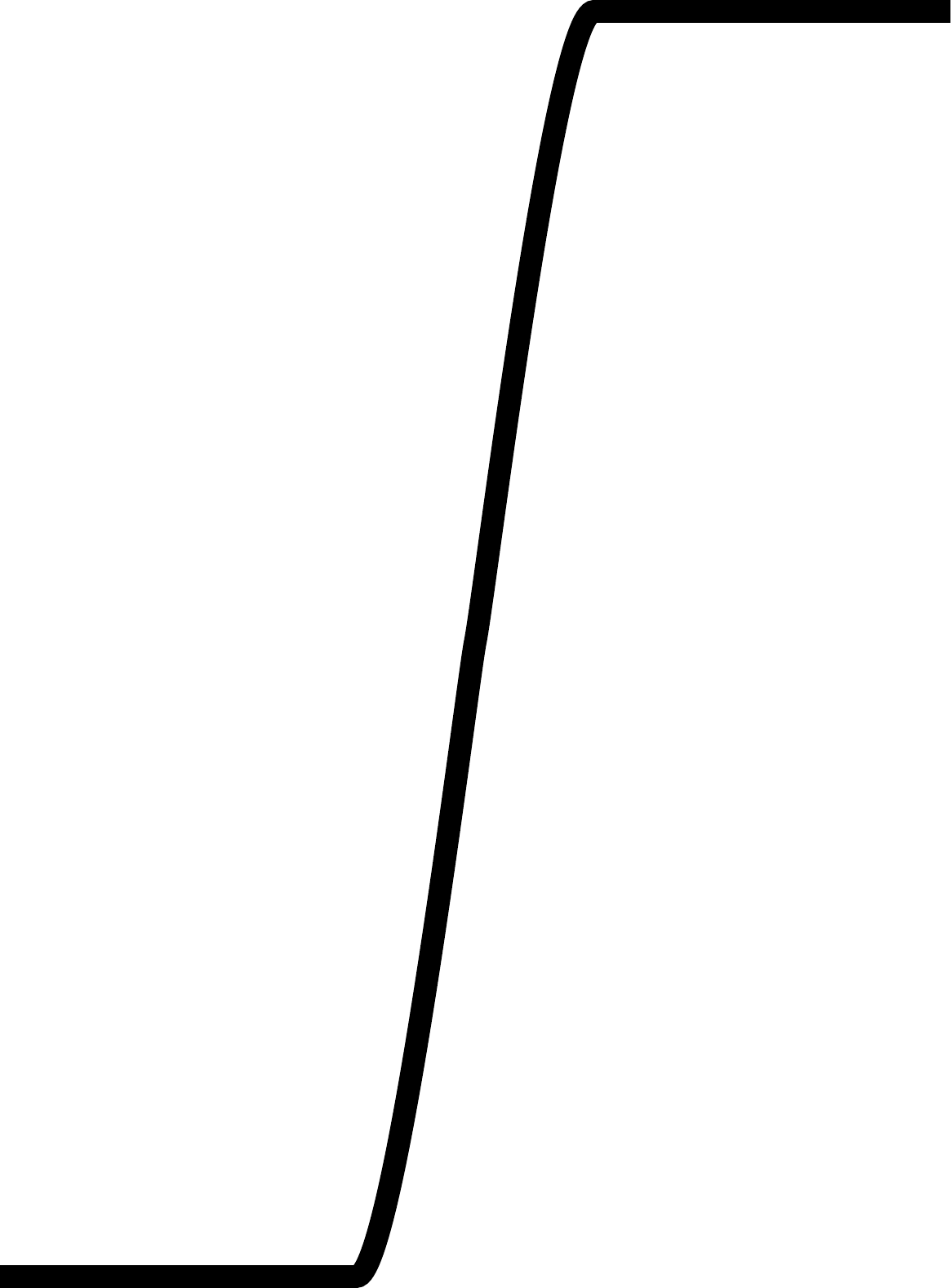}\end{array} & \rightarrow & \text{jump $\theta(n)$} & \begin{array}{c}\includegraphics[scale=0.1]{jump}\end{array}
\end{array}
\ee
Note that the peak has a magnitude $\pi$, while the height of the step is $2\pi$, and both of the contribute $+\pi$ to $\theta$ (similarly for the negative peaks, steps and jumps). If $\vartheta(n)$ only consisted of steps than one would find $\vartheta=2\theta$, but due to additional winding information contained in $\vartheta$, values of $\vartheta$ and $\theta$ are independent.


\begin{figure}[t]
\begin{minipage}{0.45\linewidth}
\includegraphics[width=\linewidth]{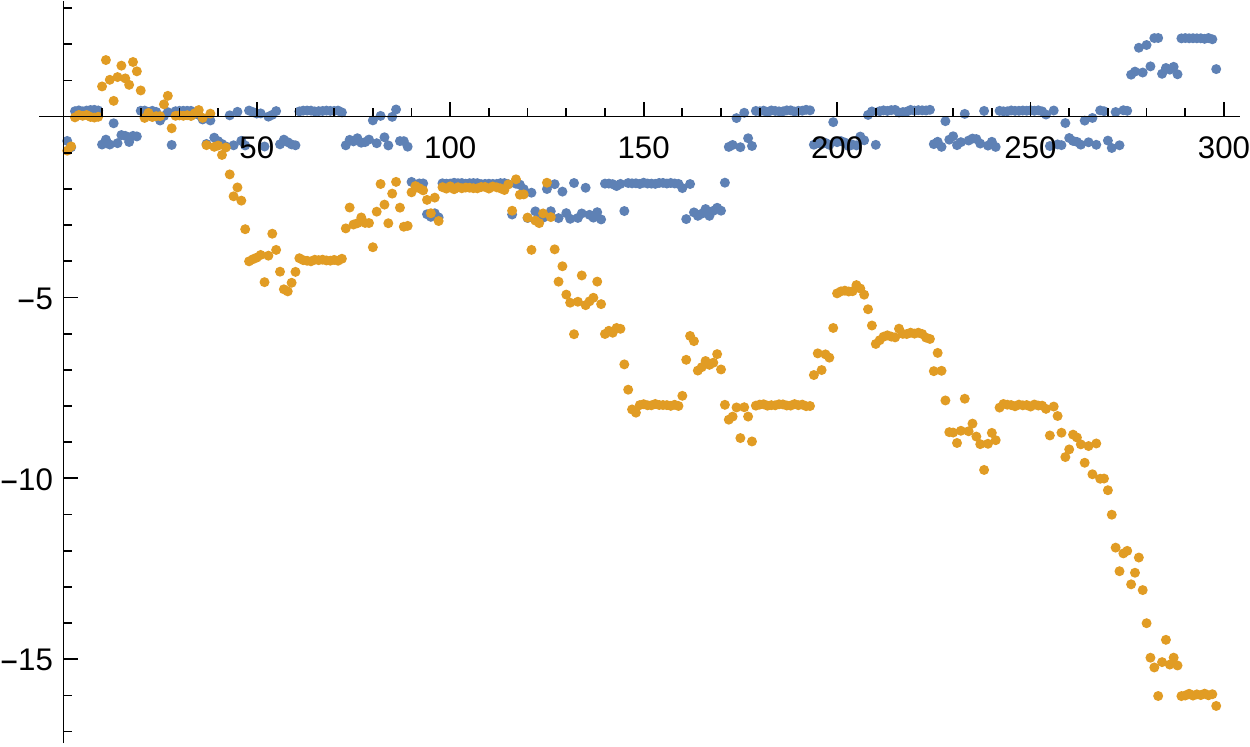}
\end{minipage}
\hfill{
\begin{minipage}{0.45\linewidth}
\includegraphics[width=\linewidth]{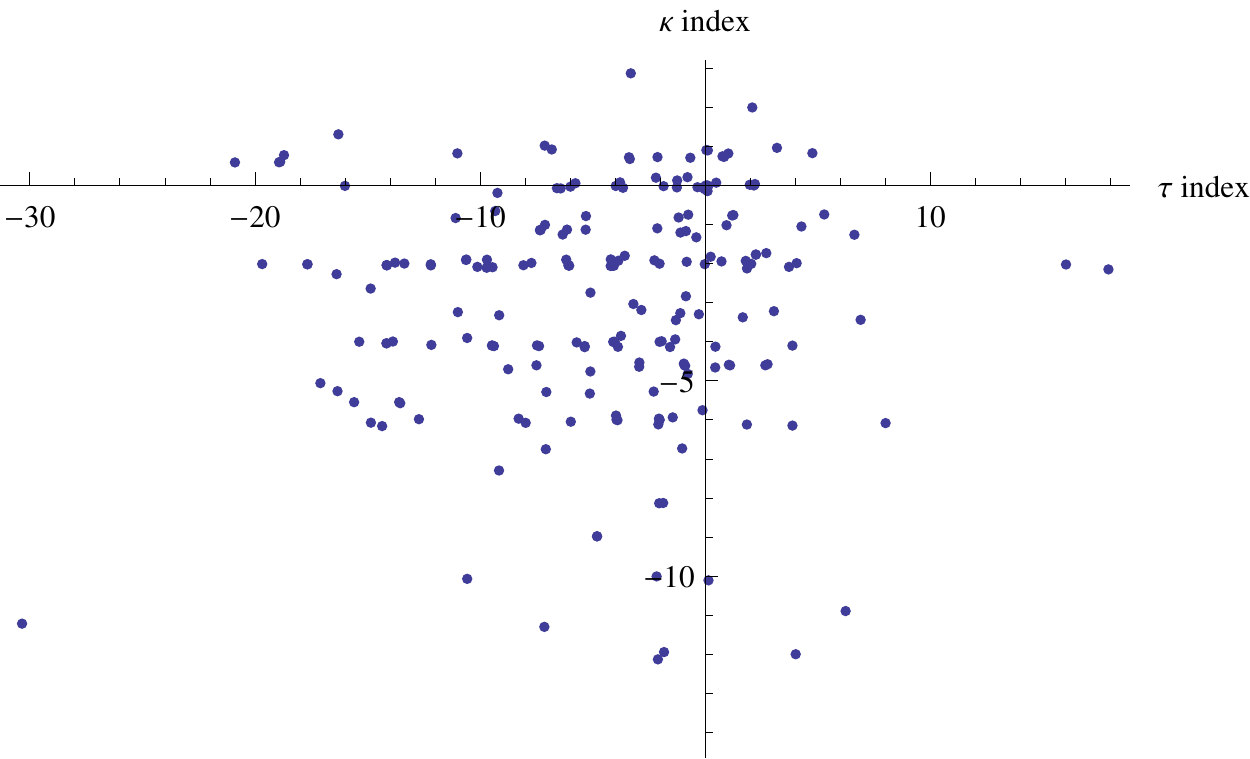}
\end{minipage}
}
\caption{(Left) Evolution of index $\vartheta$ (yellow) along the chain of xylanase ({\sl 1i1w}). The same for the $\varpi$ index (blue). (Right) Statistics of pairs $(\vartheta,\varpi)$ for 212 analyzed proteins.}  
\label{fig:1i1w}
\end{figure}


\subsection{Writhe}

Finally, we should explain the meaning of index $\varpi$. To find a non-trivial index in terms of the curvature angles we have extended their domain to negative values. As was mentioned, a positive $\kappa_n$ rotation around vector  $\bb_n$ is equivalent to a negative rotation around vector $\ub_n=-\bb_n$. So a non-trivial index $\varpi$ corresponds to a different, non-Frenet framing of the polygon. This new framing can be introduced by additional local rotations of the frame
\be
S(N) \ = \ \cdots \hat{S}_{n+1,n}\hat{S}_{n,n+1}\cdots \ = \ \cdots \hat{S}_{n+1,n}U_n^{-1}U_n\hat{S}_{n,n+1}\cdots = \cdots \hat{S}'_{n+1,n}\hat{S}'_{n,n+1}\cdots\,,
\ee
where $U_n$ rotates $\bb$ and $\nb$ vectors by $\pi$ at positions $n$. The tangent vectors remain the same, so only the framing is changed. In the new framing we will use labels $\ub_n$ for the counterparts of the binormal vectors. This rearrangement is equivalent to a gauge transformation in the definition of the self-linking number~(\ref{DiscretePexp}), so it does not change the topology.

In the $\ub$-framing we cast the rotation matrices in the form
\be
\label{uframing}
\hat{S}'_{n+1,n} \ = \  S_{n+1,n} 
\cdot {\rm e}^{i\omega_n\pi \hat{u}_n/2}\,,
\ee   
where now $\omega_n$, which can be either $0$, $1$, or $-1$, represent additional shifts of rotations relative to the uniform curvature angles. In this way, we can still assume $0\leq \kappa_n<\pi$, while the ``negative'' rotations around $u_n$ would correspond to $\omega_n=\pm 1$. This change is reflected in the transition from the curvature angle diagram on figure~\ref{fig:1a6m} (right) to the one on figure~{\ref{fig:indcor}} (right). We note that in this case there are no extra shifts in the torsion angles, since those shifts are ``undone'' by transformations $U_{n}$.

We can play the same game defining an index 
\be
\omega \ = \ \omega_1 + \ldots + \omega_{N-3}\,, 
\ee
which corresponds to collecting all the extra rotations localized at the initial frame.
\be
S(N) \ = \  {\rm e}^{i \frac{\pi}{2}\omega}\cdot S_{N-2,N-3}\cdots S_{3,2}\cdot S_{2,1}\cdot \hat{b}_1^{\omega}\,,
\ee
where, by definition, $\ub_1=\bb_1$.

Once again, imagining a closed polygon, the product of uniform rotations and $\omega\pi$ rotation around $\bb_1$ should conspire to produce a phase $\exp(i\pi\cdot Lk)$ with an integer $Lk$. As rotations around a $\bb$ vector, $\omega_n$ will contribute to the ``writhe'' part of the self-linking number.

As in the case of $\theta$, one can compare index $\omega$ with index $\varpi$ computed in section~\ref{sec:indices}. The relation between $\varpi$ and $\omega$ is the same as between $\vartheta$ and $\theta$. Consequently, one can recover $\omega$ from the evolution of $\varpi$, as in the examples of figures~\ref{fig:1a6mindex} (left) and~\ref{fig:1i1w} (left), using the rules outlined by equation~(\ref{interp}).

An obvious question is whether one should expect $\theta$ equal to $\omega$, since after all, both indices count the points with a large flip of the $\bb$ vector. It is clear from figure~\ref{fig:1i1w} (right), that $\vartheta\neq \varpi$. Moreover, one can also find examples in figures~\ref{fig:1a6mindex} (left) and~\ref{fig:1i1w} (left), where the direction of the change of $\vartheta$ is different from the direction of the change of $\varpi$. Therefore, in general, $\theta\neq\omega$. The discrepancy has the following explanation.

Indices $\theta$ and $\omega$ identify and quantify inflection points, where $\bb$ and $\nb$ have discontinuous $\pi$ jumps. The direction of the jump, or the sign of $\pi$ is not defined. Meanwhile, in the discrete polygon, like the protein molecule, there are no well-defined inflection points. What one has is some discretized resolution of the framing across inflections. The resolution depends, among other things, on independent``random fluctuations" of the vectors $\kappa_n$ and $\tau_n$ around their mean values. For example, the jump in $\theta$ is defined as $\pi\theta_n=\tau_{n}-\tau_{n-1}$, where $\tau_{n-1}$ is supposed to be close to $\bar\tau$. The difference is never exactly $\pi$, which allows to determine the sign of the rotation, but the value of the sign depends on particular values of the angles at position $n$. Consequently, the $\bb$ and $\ub$ framings represent two different topological classes characterized by two self-linking numbers, whose values differ by $|\theta-\omega|$.

Nevertheless some ``invariant" information can be obtained. Since by construction, and by observation of figures~\ref{fig:1a6mindex} (left), large jumps occur simultaneously in curvature and torsion channels, one at least expects that
\be
\theta \ = \ \omega \mod 2\,.
\ee
We checked that for $80\%$ of the analyzed proteins, rounded values of $\vartheta$ and $\varpi$ are indeed equal modulo 2. Moreover, one can count the total number of jumps, which should agree between two channels,
\be
\sum\limits_n|\omega_n| \ = \ \sum\limits_m|\theta_m|\,.
\ee
This is a topological invariant since any local large rotation of the frame removing $\theta_n$ will create a contribution to $\omega_n$.

More generally, one can define the set of vectors $\ub_n$ in terms of two-dimensional planes $V$ in the three-dimensional space. In the case of a generic plane all vectors $\bb$ will either have a positive or a negative scalar product with one of the two normal vectors to the plane. We will flip the direction of those $\bb_n \to \ub_n$ whose scalar product sign would be opposite to that of vector $\bb_1$. If $\bb_n$ and $\bb_{n+1}$ had opposite orientations (not within $V$) one of them would always become a $\ub$-vector giving a non-zero $\omega_{n+1}$. Since large rotations of $\bb$ vectors do not always happen through exactly a $\pi$ angle, there will also be non-zero $\theta$ contributions. The topological invariant of the polygon will be the sum
\be
\sum\limits_n|\omega_n| \ + \ \sum\limits_m|\theta_m|\,,
\ee
where only the values $\omega_n$ and $\theta_m$ are $V$-dependent, but not the sum of their absolute values. This is a special version of the Calugareanu's theorem.

\section{Gauge theory description}
\label{sec:gauge}

In the discrete approach three-dimensional curves (polygons) can be described using set of angles $\kappa_n$ and $\tau_n$. One might have noticed that curvature angles are associated to the nodes of the polygons, while torsion angles correspond to the bonds connecting the nodes. In the language of lattice gauge theory $\kappa_n$ can be understood as vertex (matter) degrees of freedom, while $\tau_n$ are ``connections'' (gauge fields). Indeed this point can be further elaborated in the continuous presentation (see~\cite{Danielsson:2009qm}).

In the Frenet picture curvature $\kappa(s)$ and torsion $\tau(s)$ characterize the rotation of the Frenet frame around the curve. However, the particular choice of the Frenet framing is not physical. There is an infinite number of possible choices of framing for a smooth generic curve. For example, a different local choice can be realized by introducing a local rotation of the Frenet frame around the tangent vector. It is not hard to see, that from the point of view of the Frenet description, such a local rotation induces transformations of the curvature and torsion~\cite{Danielsson:2009qm},
\be
\label{gauge}
\kappa(s) \ \to {\rm e}^{i\alpha(s)}\kappa(s)\,, \qquad \tau(s) \ \to \ \tau(s) + \alpha'(s)\,,
\ee
where it is convenient to generalize curvature to a complex quantity. Indeed, this means that curvature behaves like a complex scalar field, while $\tau$ is a one-dimensional analog of a gauge fields. 

At the same time, we have seen from equation~(\ref{Pexponent}) that one-dimensional gauge field $\tau$ and complex field $\kappa$ can be viewed as components of a non-Abelian $SU(2)$ gauge connection. Moreover the non-Abelian one-dimensional connection is a projection of a three-dimensional connection on the given curve, see \emph{e.g.}~\cite{Gordeli:2015aya}. The three-dimensional connection in turn, defines framing at any point in space. It is well known, that framed three-dimensional manifolds are classified by an integer framing number. Consequently, one can think of the framing as coming from an embedding of the curve in a framed manifold. The choice of the framed manifold will define the framing and the associated integer number -- the self-linking. Projection of the three-dimensional connection on a selected curve breaks the $SU(2)$ symmetry down to the $U(1)$ subgroup~(\ref{gauge}) of frame rotations around the tangent vector.

Based on the standard symmetry approach to construction of effective actions in field theory one can ask, what is the effective description of curves defined by functions $\kappa(s)$ and $\tau(s)$, subject to gauge transformations~(\ref{gauge}). A simplest effective model of such degrees of freedom is provided by the Abelian Higgs model~\cite{Danielsson:2009qm}. In order for the theory to reproduce solutions with non-zero torsion the minimal extension is to add a one-dimensional Chern-Simons term (see~\cite{Hu:2013,Gordeli:2015aya} for the general approach to the construction of the effective model), as in the following functional, \emph{cf.}~\cite{Danielsson:2009qm},
\be
\label{AHaction1dCS}
F \ = \  \int ds\ \left(\frac12\,(\nabla-i\htau)\hk^\ast (\nabla+i\htau)\hk   - \frac{\mu^2}2\, \left|\hk\right|^2  + \frac{\lambda}2\,|\hk|^4\right)  - F\int d s\ \htau\,,
\ee
Here hats upon torsion and curvature indicates that they are gauge dependent quantities transforming according to equations~(\ref{gauge}). In the Frenet gauge curvature is a real scalar $\hk=\kappa$ and $\htau=\tau$. Constants $\mu$, $\lambda$ and $F$ are phenomenological parameters of the curves, which are obtained by comparing the curves and basic geometric features of the protein molecules. $\nabla$ is the one-dimensional gradient. The last term is an analog of the Chern-Simons action for gauge fields in three dimensions. To make the theory regular it might also be useful to add a gauge invariant Proca mass term for the gauge field proportional to $\tau^2$.

Functional~(\ref{AHaction1dCS}) can serve as an effective free energy functional, whose minimum energy configurations are curves with constant curvature and torsion, \emph{i.e.} helices. Apart from the lowest energy translationally invariant solutions (ground states) such theories can also have solitons, \emph{i.e.} solutions interpolating between the same, or different ground states~\cite{Chernodub:2010xz}. The solitons can be either kinks or lumps (dents) in curvature and torsion.  Clearly plots of the topological indices on figures~\ref{fig:1a6m} and~\ref{fig:1i1w} are examples of such solitons. In other words, topological indices $\theta$ and $\omega$ count the total soliton numbers of the proteins. We do not intend to describe the corresponding solitons here, though many details about their discrete version, in direct application to proteins, can be found in works~\cite{Molkenthin:2011,Hu:2011,Krokhotin:2012,Krokhotin:2012b}.


\section{Conclusions}
\label{sec:conclusions}

In this paper we considered effects of framing for nature given discretizations of smooth curves, such as protein molecules. Protein backbones equipped with the Frenet framing underline basic features of the secondary structure of proteins, such as existence of regular helical pieces connected by structural kinks (motifs). Studying the structure at the topological level we have introduced a simple spin-chain-like model of proteins, in which regular pieces corresponded to parallel oriented vectors (ground state), while kinks corresponded to short sequences of vectors that deviated from the standard orientation.

We have demonstrated that the kinks-motifs can be counted by topological indices taking discrete values. This owes to the fact that the kinks correspond to inflection points of smooth versions of the discrete curves, where the Frenet frame experiences large rotations. The discrete indices  can be related to the self-linking number $Lk$ of a framed curve, which is a topological invariant. Although the self-linking number is defined for smooth framed curves, we showed how the notion of the self-linking number can be extended to the case of singular framings, such as the Frenet framing with inflection points. In this case the invariant splits into a sum of the regular self-linking number of a regular framing and an integer index associated to the inflection points.

We have shown that the Frenet framing corresponds to a special type of the singular self-linking number -- a pure twist. To have also a writhe-type contributions to $Lk$ one has to flip at least a part of the normal vectors in the Frenet framing. Such flips are local (gauge) transformations that convert units of twist to units of writhe in accordance with the Calugareanu's theorem. We showed that in terms of the indices defined here, the theorem is satisfied only if we compute all the indices modulo sign. Otherwise our indices are defined with respect to incompatible framings and  cannot be compared. This is, of course, a result of the singularity of the Frenet framing.

Spin chains can be extended to effective field theory models. In such models curvature is promoted to a complex scalar field $\kappa(s)$, while torsion is a one-dimensional scalar field. Indices computed here can be related to topological charges of field theory solitons.

It is interesting to study how the topological indices defined here behave in the protein's dynamics. Formation of the structural motifs is one of the basic steps in the protein folding. We are offering a slightly different point of view on this process: it would be interesting to study the evolution of the folding indices during the folding process.

\paragraph{Acknowledgements}  DM and AJN would like to thank the participants of the workshop ``Physics and Biology of Proteins'' held at the International Institute of Physics in Natal in June 2017 for useful discussions. The work of DM and AS was supported by the binational Armenian-Russian collaboration through the grant No~RFBR-18-51-05015~Arm of Russian Foundation of Basic Research and the grant No 18RF-039 of the Armenian Research Council (ARC). The work of AS was also partially supported by the ARC grant 18T-1C153 and by the Brazilian Ministry of Education. The work of AJN was supported by Qian Ren program at the Beijing Institute of Technology, by a grant 2018-04411 from Vetenskapsr{\aa}det, by the Carl Trygger Stiftelse and by the Henrik Granholms Stiftelse. AJN also acknowledges {\it  Eutopia} (EU-COST Action CA17139). In the analysis of the PDB data the authors have used the list of high-resolution protein structures provided by K.~Hinsen.  Additionally DM would like to thank the hospitality of Nordita during his visit to Stockholm in June and July 2018.

\end{document}